\documentstyle[prl,preprint,aps,epsf]{revtex} 

\newcommand{\postscript}[2]{
\setlength{\epsfxsize}{#2}
\centerline{\epsfbox{#1}}}           
\newlength{\headroom}
\newlength{\psfigskip}
\begin{document}
%\twocolumn
% *** Title Page **************************************************************
\title{
		     Solutions to the Solar Neutrino Anomaly
}
\author{             
		     Naoya Hata$^1$ and Paul Langacker$^2$
}

\address{
{\it
		     $^1$ Institute for Advanced Study 
	             Princeton, NJ 08540
}}
\address{
{\it
                     $^2$Department of Physics,                       
                     University of Pennsylvania,               \\              
                     Philadelphia, Pennsylvania 19104          \\ 
}}

                     \date{ 
%                            \today,
                             May 16, 1997,
                             IASSNS-AST 97/29, UPR-751T}

\maketitle
%
% *** Abstract ****************************************************************

%\vspace*{-2.0cm}
%\renewcommand{\baselinestretch}{1.3}
\renewcommand{\baselinestretch}{0.95}
\begin{abstract}

We present an updated analysis of astrophysical solutions, two-flavor
MSW solutions, and vacuum oscillation solutions to the solar neutrino
anomaly.  The recent results of each of the five solar neutrino
experiments are incorporated, including both the zenith angle
(day-night) and spectral information from the Kamiokande experiment,
and the preliminary Super-Kamiokande results.  New theoretical
developments include the use of the most recent Bahcall-Pinsonneault
flux predictions (and uncertainties) and density and production
profiles, the radiative corrections to the neutrino-electron
scattering cross section, and new constraints on the Ga absorption
cross section inferred from the gallium source experiments.  From a
model independent analysis, arbitrary astrophysical solutions are
excluded at more that 98\% C.L.  even if one ignores any one of the
three classes of experiment, relaxes the luminosity constraint, or
allows more suppression of the $^7$Be than $^8$B flux.  The data is
well described by large and small mixing angle two-flavor MSW
conversions, MSW conversions into a sterile neutrino with small
mixing, or vacuum oscillations.  We also present MSW fits for
nonstandard solar models parameterized by an arbitrary solar core
temperature or arbitrary $^8$B flux.

\end{abstract}
\pacs{PACS numbers: }
%
% *** Text ********************************************************************

%\twocolumn
\renewcommand{\baselinestretch}{1.00}

%    		*	*	*	*	*	*  

\section{Introduction}
\label{Sec:Introduction}

The solar neutrino problem currently provides one of the most
compelling experimental signatures for the physics beyond the standard
model.  The Mikhyev-Smirnov-Wolfenstein (MSW) effect \cite{MSW} via
neutrino mass and mixing provides a complete explanation of the
existing solar neutrino data, while astrophysical solutions, even
those with drastic alterations of the standard solar model, simply
fail.  The difficulty with astrophysical explanations persists even if
we ignore data of any one of the three kinds of experiments, {\it
i.e.}, the Homestake chlorine experiment \cite{Homestake}, the water
\v{C}erenkov experiments of Kamiokande \cite{Kamiokande} and 
Super-Kamiokande \cite{Super-Kamiokande}, or the gallium experiments
of SAGE \cite{SAGE} and GALLEX \cite{GALLEX}.  (The experimental
results are summarized in Table~\ref{Tab:Data} along with the SSM
predictions \cite{Bahcall-Pinsonneault-95}.)  The successful
results of gallium source experiments
\cite{GALLEX,SAGE} and the excellent agreement between
the standard solar model predictions and the recent helioseismology
data further reinforce our confidence in neutrino oscillation
solutions \cite{Bahcall-Pinsonneault-Basu-Christensen-Dalsgaard}.  The
new generation of solar neutrino experiments, such as Super-Kamiokande
and Sudbury Neutrino Observatory (SNO), will provide critical tests of
the MSW predictions for the neutrino energy spectrum distortion, the
day-night rate asymmetry, and the charged-to-neutral current ratio.

In this paper we examine the current status of the solar neutrino
problem for astrophysical solutions, MSW solutions, and vacuum
oscillation solutions.  The data as of February 1997, including the
final Kamiokande results and the preliminary Super-Kamiokande data,
are used.  This is the first MSW analysis using the entire data of the
Kamiokande spectrum and day-night asymmetry.  We also incorporate the
latest standard solar model with diffusion effects
\cite{Bahcall-Pinsonneault-95}, the radiative corrections for the
neutrino-electron scattering cross section
\cite{Bahcall-Kamionkowsky-Sirlin}, the improved determination of the
$^8$B decay spectrum \cite{Bahcall-etal-96}, and the constraint on the
gallium cross section from the source experiments \cite{Hata-Haxton}.
The calculations in this paper are described in detail in our previous
works
\cite{BHKL,Hata-Bludman-Langacker,Hata-Langacker-93,Hata-Langacker-94,%
Hata-Langacker-95}, including the model-independent analysis for
astrophysical solutions, MSW calculations, the day-night effect, and
consistent treatment of solar model uncertainties.  We consider only
two-flavor oscillations because of their simplicity and viability.  We
referred to Ref.~\cite{Fogli-Lisi} for a recent analysis for
three-flavor oscillations and Ref.~\cite{Smirnov-96} for recent
developments in neutrino physics.

This paper is organized as follows.  In
Section~\ref{Sec:Astrophysical-Solutions} we reexamine the general
astrophysical solutions and show their failure with much stronger
statistical significance than before.  This is true even if we ignore
any one of the three types of experiment or the solar luminosity
constraint.  We also discuss Monte Carlo evaluations of goodness of
fit when the number of degrees of freedom (DOF) effectively becomes
zero or negative.  In Section~\ref{Sec:MSW-Solutions} the constraints
on the MSW parameters are updated.  The Kamiokande spectrum result by
itself excludes the adiabatic (horizontal) branch almost entirely.
The MSW solutions with nonstandard core temperatures, and with
nonstandard $^8$B flux, and oscillations to sterile neutrinos are also
examined.  Vacuum oscillation solutions are discussed in
Section~\ref{Sec:Vacuum-Oscillation-Solutions}.  We show that the
Kamiokande spectrum data considerably restricts the allowed parameter
space.  The conclusions of our analysis are given in
Section~\ref{Sec:Conclusions}.

%    		*	*	*	*	*	*  

\section{Astrophysical Solutions}
\label{Sec:Astrophysical-Solutions}

The incompatibility of astrophysical solutions and the solar neutrino
data has been investigated in many ways.  These include the failure of
explicit nonstandard solar models \cite{Bahcall-Ulrich,Bahcall-book},
the comparison of the Homestake and Kamiokande results
\cite{Bahcall-Bethe}, lower core temperature fits
\cite{Bludman-Kennedy-Langacker,BHKL},
and so on.  One can generalize the argument against astrophysical
solutions by a model independent analysis using $pp$, $^7$Be, $^8$B,
and CNO fluxes as free parameters under minimal assumptions on the
solar luminosity, the beta spectrum shape, and the detector cross
sections.  The details of our analysis is described in
\cite{Hata-Bludman-Langacker,Hata-Langacker-95} (similar analyses are 
found in \cite{Spiro-Vignaud,Castellani-Degl'Innocenti-Fiorentini-AA,%
Castellani-Degl'Innocenti-Fiorentini-PRD,Parke}).  We will display the
results of the fits in the $\phi({\rm Be})- \phi({\rm B})$ plane, both
normalized to the SSM values ($\phi({\rm Be})_{\rm SSM} = 5.15 \times
10^9$ and $\phi({\rm B})_{\rm SSM} = 6.62 \times 10^6$ in units of
cm$^{-2}$s$^{-1}$ \cite{Bahcall-Pinsonneault-95}).

The constraints from individual data are shown in
Fig.~\ref{Fig:BeB_each}.  The combined result from Kamiokande and
Super-Kamiokande determines the $^8$B flux only.  The CNO flux as well
as the $^7$Be and $^8$B fluxes are used as free parameters in fitting
the Homestake result.  In fitting to the combined Ga result from SAGE
and GALLEX, the $pp$ flux is also varied as a free parameter subject
to the luminosity constraint.

A comparison of Fig.~\ref{Fig:BeB_each} with our original analysis in
1993 (Fig.~1 in Ref.~\cite{Hata-Bludman-Langacker}) displays a
dramatic improvement in statistics, especially in the water
\v{C}erenkov data and the gallium data.  The addition of
the high-statistics Super-Kamiokande result with about 1000 events from
the first 100 days
\footnote{The previous Kamiokande experiment collected a total of 
only 600 events in 5.7 years.}
has reduced the uncertainty in the $^8$B flux measurement in half.
The low rate and the precision of the gallium result alone impose
serious problems for astrophysical solutions.  The $^8$B flux allowed
by the Homestake and gallium data each is smaller than the
Kamiokande--Super-Kamiokande measurement for almost the entire range
of the $^7$Be flux.  In addition the Homestake and gallium together
are incompatible, since for $\phi({\rm Be}) \sim 0$ and $\phi({\rm B})
\sim 0$, the gallium data requires the CNO flux to be $\sim 0$, while
the Homestake data requires the CNO flux 4.9 times larger than the SSM
value.

The severity of the problem with astrophysical solutions can be seen
by applying the joint analysis to all the data, shown in
Fig.~\ref{Fig:BeB_comb_cnozero}.  We allow the $^7$Be flux to
be negative.
\footnote{We can also allow the CNO flux to be negative.  In this
case the allowed region is essentially unbounded for $\phi({\rm Be})
\ll 0$.}
The best fit of the combined observations is in the non-physical
region:
$\phi({\rm Be})/\phi({\rm Be})_{\rm SSM} = - 0.6 \pm 0.4$ and 
$\phi({\rm B}) /\phi({\rm B})_{\rm SSM}  =   0.4 \pm 0.05$
with $\chi^2_{\rm min} = 0.5$ for 3 data points, one luminosity
constraint, and four free parameters: i.e., zero degrees of freedom
(DOF).  Within physical parameter space ($\phi({\rm Be}) \geq 0$ and
$\phi({\rm B}) \geq 0$), the best fit is
$\phi({\rm Be})/\phi({\rm Be})_{\rm SSM} < 0.1$ and 
$\phi({\rm B}) /\phi({\rm B})_{\rm SSM}  =   0.38 \pm 0.05$
with $\chi^2_{\rm min} = 9.2$.  The usual prescription for goodness of
fit (GOF) evaluations \cite{Numerical-Recipes} does not apply since we
have zero DOF. (Later we will encounter fits with negative DOF).
In addition the probability distribution is non-Gaussian due to the
physical constraints, {\it i.e.}, the fluxes should be non-negative.
Generalization of GOF by employing the Monte Carlo method is
necessary.

GOF in this case is defined by the probability to obtain $\chi^2_{\rm
min}$ as large as 9.2 or larger by chance due to the experimental
uncertainties, if the best fit fluxes are true.  Our Monte Carlo
construction is (1) to take the central flux values of the fit, (2)
calculate the solar neutrino rates for the three experiments, (3)
generate Monte Carlo distributions for each experiment assuming the
actual experimental uncertainties (7.7, 7.4, and 9.7\% for the
Homestake, Kamiokande/Super-Kamiokande, and gallium experiments,
respectively), (4) for each Monte Carlo data set, apply our model
independent analysis to obtain the $\chi^2$ minimum.  Note that when
one has $N$ parameters and $M$ constraints with Gaussian errors and $N >
M$, this procedure can be done analytically, reproducing the
usual $\chi^2$ distribution for $N - M$ dimensions \cite{Fermi-note}.

The Monte Carlo distribution of the $\chi^2$ minima is shown in
Fig.~\ref{Fig:gof_lin_comb}; the $\chi^2_{\rm min} = 9.2$ from the
actual data is also indicated.  The probability of getting $\chi^2$
minimum larger than 9.2 by chance is 0.6\%.  That is, our model
independent analysis excludes the best fit astrophysical solution
at the 99.4\% C.L.  
\footnote{
We have also considered two alternative nonstandard evaluations of
GOF.  If one assumes (somewhat arbitrarily) that the CNO flux is fixed
to zero \cite{Hata-Bludman-Langacker}, the fit is for 1 DOF and
$\chi^2_{\rm min} = 9.2$ corresponds to 99.8\% C.L.  If one defines
GOF by the ratio of the volume of the likelihood function integrated
within the physical region to the volume integrated in the entire
parameter space (including negative fluxes), the ratio is 0.2\%, or
the exclusion is 99.8\% C.L.  Thus, the both estimates are similar to
the Monte Carlo result.}

Next we consider the same analysis but ignoring one of the
constraints.  Fig.~\ref{Fig:BeB_kam+ga}, \ref{Fig:BeB_cl+ga}, and
\ref{Fig:BeB_kam+cl}  show the results each without the Homestake data,
the water \v{C}erenkov data, and the gallium data, respectively.
Fig.~\ref{Fig:BeB_comb_nolumi} is the result with all experiments, but
without the luminosity constraint.  (Violations of the luminosity
constraint would be possible if the properties of the solar core were
somehow varying on a time scale short compared with $10^4$ years.)
The corresponding GOF of the best fit in the physical region is 98.9,
98.3, 98.9, and 98.3\% C.L., respectively.  Although the constraints
on the fluxes are somewhat relaxed by ignoring one of the data or the
luminosity constraint, the essential problem with the poor fit
remains.  In addition one can also see the persistent problem of the
strong suppression of the $^7$Be flux, which is difficult to obtain by
astrophysical effects in general.

\section{MSW Solutions}
\label{Sec:MSW-Solutions}

\subsection{Kamiokande spectrum and day-night data}

The Kamiokande experiment has completed its measurements and published
the results of total rate, spectrum data, and day-night data
\cite{Kamiokande}.  The MSW parameters can be constrained by those 
results.  Note that the spectrum distortions and day-night
time-dependence are not expected with standard neutrino physics and
are powerful indicators of physics beyond the standard model.
The spectrum shape measured in Kamiokande is consistent with the one
expected from the undistorted $^8$B $\beta$-decay spectrum albeit the
uncertainties are large.  The shape is inconsistent with the strong
suppressions at large energies expected in the MSW adiabatic branch
($\Delta m^2 \sim 10^{-4}$ eV$^2$ and $\sin^22\theta
\sim 10^{-4} - 0.1$), and the data exclude the region almost entirely
as shown in Fig.~\ref{Fig:k_exclusion}.  This exclusion is simply due
to lack of distortions in the data and independent of the
uncertainties in the initial $^8$B flux.  The data, however, do not
constrain the nonadiabatic (diagonal) branch, in which spectrum
distortions are smaller.

The day-night result was published as one day-time rate and five bins
for night-time bins.  The binning was $\cos \theta = 0 - 0.2$, $0.2 -
0.4$, ... $0.8 - 1$, where $\theta$ is the angle between the direction
to the Sun and the nadir at the detector.  Within the experimental
uncertainties the six bins are consistent and thus exclude a large
region in which day-night asymmetries due to the Earth effect are
expected.  The excluded region is shown in
Fig.~\ref{Fig:k_exclusion}.

The allowed parameter space from the total Kamiokande rate is shown in
Fig.~\ref{Fig:fkam_ave}.  This constraint is model dependent and
we have assumed the Bahcall-Pinsonneault model (1995)
\cite{Bahcall-Pinsonneault-95} including its
uncertainties.  In this and other fits the correlation in the
theoretical uncertainties between the flux components and between
the experiments are included.  

Unfortunately we cannot combine the spectrum and day-night data since
the errors in those are strongly correlated and the correlation matrix
is unpublished.  Fig.~\ref{Fig:fkam_sp} and \ref{Fig:fkam_dn} show the
allowed region when the total rate is combined with the spectrum data
and day-night data, respectively.

\subsection{Preliminary results from Super-Kamiokande}

Recently the Super-Kamiokande collaboration reported the results of
about 1000 events from the first 100 days of data
\cite{Super-Kamiokande}.  The total rate (see Table~\ref{Tab:Data}) 
is consistent with the previous Kamiokande rate, and new uncertainties
are much smaller.  When combined with the Kamiokande total rate, the
error is reduced almost by half.  The new constraint is shown in
Fig.~\ref{Fig:fkam_comb}.  The uncertainties in the MSW parameter
space are now dominated by the $^8$B flux error in the solar model
calculations ($\sim$ 15\%).  Future measurements of model-independent
quantities such as the spectrum shape and day-night effect and also
the charged-to-neutral current ratio in SNO, are essential to confirm
the MSW interpretation and to improve the determination of the MSW
parameters.

\subsection{MSW combined results}

We next consider the MSW constraint including the Homestake and
gallium data.  The two separate allowed regions are shown in
Fig.~\ref{Fig:p_comb}.  The fit includes Kamiokande day-night data and
the averaged Super-Kamiokande data.  (The allowed regions are
essentially identical even if the Kamiokande spectrum data are used.)
Both allowed regions provide a good fit.  The $\chi^2$ minimum for 7
degrees of freedom is 5.9 (55\% C.L.) and 6.4 (49\% C.L.) for the
small-angle and large-angle solution, respectively.  (Details are
listed in Table~\ref{Tab:MSW-fit}.)  The fit for the large-angle
solution improved from previous analyses (see for example
Ref~\cite{Hata-Langacker-94}) due to the larger $^8$B flux in the new
SSM and a new, smaller Kamiokande and Super-Kamiokande rate, both of
which reduces the relative difference between the Homestake rate and
the Kamiokande/Super-Kamiokande rate and allows energy-independent
flux reduction as expected in the large-angle region.

We have also considered oscillations to sterile neutrinos
\cite{sterile-nu}.  The GOF for the large angle region is
94\% C.L.\ However, the 95\% allowed region defined by 
$\chi^2 < \chi^2_{\rm min} + 6.0$ does not appear in the
$\sin^22\theta$ -- $\Delta m^2$ parameter space
(Fig.~\ref{Fig:p_comb_sterile} and Table~\ref{Tab:MSW-fit-sterile}).
The large angle solution for sterile neutrinos is also severely
constrained by big bang nucleosynthesis \cite{sterile-nu,Hata-etal-95}.
 
While nonstandard solar models, as discussed in
Section~\ref{Sec:Astrophysical-Solutions}, cannot solve the solar
neutrino problem, the MSW effect can be also considered with
nonstandard solar models \cite{BHKL,Hata-Langacker-94}.  Many of those
models may be parameterize by nonstandard core temperature ($T_C$) or
simply a nonstandard $^8$B flux, whose uncertainties might be larger
than the SSM estimate.  We consider joint fits of $T_C$ or $^8$B flux,
in addition to the MSW parameters.

When $T_C$ is used as a free parameter, the neutrino fluxes can be
scaled according to the power law.  From the Monte Carlo investigation
of the SSM, the indices of the power law are obtained in
Ref.~\cite{Bahcall-Ulmer}, based on the Monte Carlo SSMs by Bahcall
and Ulrich \cite{Bahcall-Ulrich}. 
\footnote{
The Monte Carlo estimate for the model with diffusion is not yet available.}
The combined Homestake, gallium, Kamiokande, and Super-Kamiokande data
constrain
\begin{equation}
	T_C / T_C^{\rm SSM} = 0.99 ^{+0.02}_{-0.03} 
        \hspace{1em} (1 \sigma),
\end{equation}
and $T_C/T_C^{\rm SSM} = 0.95 - 1.02$ for 95\% C.L., where 
$T_C^{\rm SSM}$ is the the SSM value ($1.567 \times 10^7$K).  This
result is in excellent agreement with the SSM range $1 \pm 0.006$
\cite{Bahcall-Ulrich}: the data are consistent with the SSM prediction
in the presence of the MSW effect.  Our likelihood for $T_C$ is shown
in Fig.~\ref{Fig:tc-P}.  The corresponding MSW parameter space is
shown in Fig.~\ref{Fig:p_comb_tcfree}.

Next the $^8$B flux is used as a free parameter, the combined data
determines
\begin{equation}
	\phi({\rm B}) / \phi({\rm B})_{\rm SSM}
	   = 0.76 ^{+0.38}_{-0.30}  \hspace{1em} (1 \sigma).
\end{equation}
and 0.31 -- 1.50 for 95\% C.L.\  
Although the uncertainty is large, the result is consistent with
the SSM range $1 ^{+0.14}_{-0.17}$ \cite{Bahcall-Pinsonneault-95}.
Our likelihood for $\phi({\rm B})$ and the corresponding MSW regions
are shown in Fig.~\ref{Fig:B-P} and \ref{Fig:p_comb_Bfree}.

\section{Vacuum Oscillation Solutions}
\label{Sec:Vacuum-Oscillation-Solutions}

The simple two-flavor vacuum oscillations are still a
phenomenologically viable solution \cite{Krastev-Petcov}.  Those
solutions require tuning of $\Delta m^2$ and the Sun-Earth distance at
the 5\% level to explain the observations, which is a conceptual setback.

Some parameter space for the vacuum oscillation for $\Delta m^2 \sim
10^{-10}$ eV$^2$ predicts a relatively strong energy dependence, and
the recent Kamiokande spectrum data alone can exclude a wide range of
parameters, as shown in Fig.~\ref{Fig:vexclusion_kam_sp}.  When
combined with the results of Homestake, gallium, and Super-Kamiokande,
we find three separate allowed regions within a narrow range of
parameters [$\Delta m^2 = (5-8) \times 10^{-11}$ eV$^2$ and
$\sin^22\theta = 0.65 - 1$] as shown in
Fig.~\ref{Fig:vfcomb+sk+kam_sp}.  The GOF for the best fit parameters
is 9.9 for 9 DOF, which is acceptable.  Details of the fits are
listed in Table~\ref{Tab:vacuum-fit} For comparison we show five
allowed regions when the Kamiokande spectrum data are ignored
Fig.~\ref{Fig:vfcomb+sk}).

\section{Conclusions}	
\label{Sec:Conclusions}

Although the general scope of the solar neutrino problem has not
chanced since the first result of the gallium experiment in 1992, the
improved accuracy of the solar neutrino data provide a more robust
assessment of solutions.  The astrophysical solutions in general have
difficulties unless all experiments are wrong, or at least two out of
three data and the SSM are wrong.

The MSW effect provides viable solutions: the small mixing-angle
solution ($\sin^22\theta \sim 0.008$ and $5 \times 10^{-6}$ eV$^2$)
and the large-angle solution ($\sin^22\theta \sim 0.6$ and $1.6 \times
10^{-5}$ eV$^2$), assuming the latest SSM by Bahcall and Pinsonneault.
Oscillations to sterile neutrinos are possible, but only for small
angles.  When the core temperature or the $^8$B flux is used as a free
parameter, the joint data determines those at the 3\% and 30\% level,
respectively.  Those ranges are consistent with the SSM predictions.
Vacuum oscillations are still viable for $\Delta m^2 \sim 6 \times
10^{-11}$ eV$^2$ and $\sin^22\theta \sim 0.9$.

The Kamiokande day-night data and spectrum data each exclude a large
parameter space for MSW, independent of SSM predictions.  We expect
Super-Kamiokande will provide those with much improved accuracy and
eventually, along with the SNO neutral current measurement, single out
the solution of the solar neutrino problem.

\acknowledgements

It is pleasure to thank J.\ Bahcall, B.\ Balantekin, P.\ Krastev, and
A.\ Yu Smirnov for useful discussions.  This work is supported by the
National Science Foundation Contract No. NSF PHY-9513835 at the
Institute for Advanced Study and Department of Energy Contract No.DE-AC02-76-ERO-3071 at the University of Pennsylvania.

% ***  The end of text  *******************************************************
%******************************************************************************
%
%                        * * * REFERENCES * * *
%

%******************************************************************************

% Here comes the tables*** ****************************************************
\renewcommand{\baselinestretch}{1.3}

%\renewcommand{\baselinestretch}{1.0}

%                        * * * TABLES * * *

\begin{table}[hbt]
\caption{
%
%                                  TABLE I
%
The standard solar model predictions of Bahcall and Pinsonneault (BP SSM)
\protect\cite{Bahcall-Pinsonneault-95} and the results of
the solar neutrino experiments.
}
\label{Tab:Data}
\vspace{1.0ex}

%\squeezetable

\begin{tabular}{l  c c }
%
%\hline%-----------------------------------------------------------------------
%\hline%-----------------------------------------------------------------------
               & BP SSM                    & Experiments          \\[1ex]
\hline\\%[-2ex]%---------------------------------------------------------------
Homestake 
	       &  9.3 $^{+1.2}_{-1.4}$ SNU & 
                    2.55 $\pm$ 0.14 $\pm$ 0.14 SNU (0.273 $\pm$ 0.021 BP SSM)\\
\hline\\[-2ex]%----------------------------------------------------------------
Kamiokande 
	       &  & 2.80 $\pm$ 0.19 $\pm$ 0.33 \tablenotemark[1] 
                                                  (0.423 $\pm$ 0.058 BP SSM) \\
Super-Kamiokande 
	       & 6.62 $^{+0.93}_{-1.12}$ \tablenotemark[1]    & 
            2.51 $^{+0.14}_{-0.13}$ $\pm$ 0.18 \tablenotemark[1] 
                                                  ($0.379 \pm 0.034$ BP SSM) \\
Combined 
	       &  & 2.586 $\pm$ 0.195 \tablenotemark[1] 
                                                  (0.391 $\pm$ 0.029 BP SSM) \\
\hline\\%[-2ex]%---------------------------------------------------------------
SAGE                         & & 
               69 $\pm$ 10 $^{+5}_{-7}$ SNU (0.504 $\pm$ 0.089 BP SSM) \\
GALLEX      & 137 $^{+8}_{-7}$ SNU
            & 69.7 $\pm$ 6.7 $^{+3.9}_{-4.5}$ SNU (0.509 $\pm$ 0.059 BP SSM) \\
Combined    &               & 69.5 $\pm$ 6.7 SNU (0.507 $\pm$ 0.049 BP SSM) \\
%\hline%-----------------------------------------------------------------------
%\hline%-----------------------------------------------------------------------
\end{tabular}
\vspace{1ex}
\tablenotetext[1]{%
In units of $10^6$ cm$^{-2}$sec$^{-1}$.}
\end{table}

\begin{table}[hbt]
\caption{
%
%                                  TABLE II
%
The best fit parameters, the $\chi^2$ minimum, and confidence levels
of GOF for the combined MSW fits.
}
\label{Tab:MSW-fit}
\vspace{1.0ex}

%\squeezetable

\begin{tabular}{ l  c c  }
%
%------------------------------------------------------------------------------
%------------------------------------------------------------------------------
		  &	Small Angle  &  Large Angle   \\
\hline%------------------------------------------------------------------------
  $\sin^22\theta$      & $8.2\times 10^{-3}$ & 0.63      \\
  $\Delta m^2$ (eV$^2$)& $5.1\times 10^{-6}$ & $1.6 \times 10^{-5}$ \\
\hline%------------------------------------------------------------------------
  $\chi^2$ (7 d.f.) &  5.9              & 6.3           \\
  $P$ (\%)          &  45               & 49        	\\
%
%------------------------------------------------------------------------------
%------------------------------------------------------------------------------
%
\end{tabular}
\end{table}

\begin{table}[hbt]
\caption{
%
%                                  TABLE III
%
The best fit parameters, $\chi^2$ minimum, and GOF for the combined
MSW fits for oscillations to sterile neutrinos.
}
\label{Tab:MSW-fit-sterile}
\vspace{1.0ex}

%\squeezetable

\begin{tabular}{ l  c c  }
%
%------------------------------------------------------------------------------
%------------------------------------------------------------------------------
		  &	Small Angle  &  Large Angle   \\
\hline%------------------------------------------------------------------------
  $\sin^22\theta$      & $1.0\times 10^{-2}$ & 0.72      \\
  $\Delta m^2$ (eV$^2$)& $4.0\times 10^{-6}$ & $8.9 \times 10^{-6}$ \\
\hline%------------------------------------------------------------------------
  $\chi^2$ (7 d.f.) &  6.7              & 13.7          \\
  $P$ (\%)          &  54               & 94        	\\
%
%------------------------------------------------------------------------------
%------------------------------------------------------------------------------
%
\end{tabular}
\end{table}

\begin{table}[hbt]
\caption{
%
%                                  TABLE IV
%
The best fit parameters, $\chi^2$ minimum, and GOF for the combined
vacuum oscillation fits including the Kamiokande spectrum data. 
}
\label{Tab:vacuum-fit}
\vspace{1.0ex}

%\squeezetable

\begin{tabular}{ l  c c c }
%
%------------------------------------------------------------------------------
%------------------------------------------------------------------------------
		       & Solution 1      & Solution 2     & Solution 3   \\
\hline%------------------------------------------------------------------------
  $\sin^22\theta$      & 0.83                  & 0.90     & 1.0         \\
  $\Delta m^2$ (eV$^2$)& $7.9\times 10^{-11}$ & $6.6\times 10^{-11}$ 
                                               & $5.2\times 10^{-11}$   \\
\hline%------------------------------------------------------------------------
  $\chi^2$ (9 d.f.) &  11.4              & 9.9           & 11.9 \\
  $P$ (\%)          &  75                & 64        	 & 78\\
%
%------------------------------------------------------------------------------
%------------------------------------------------------------------------------
%
\end{tabular}
\end{table}

%****  Figures ****************************************************************

\onecolumn
\renewcommand{\baselinestretch}{1.0}
%\newlength{\psfigsize}
%\setlength{\psfigsize}{0.80\hsize}
\newlength{\yysize}
\setlength{\yysize}{0.90\hsize}
\newlength{\squaresize}
\setlength{\squaresize}{1.00\hsize}
\newlength{\pslhfsize}
\setlength{\pslhfsize}{1.00\hsize}
\newlength{\captionhead}
\setlength{\captionhead}{-15ex}
\newlength{\captionheadmsw}
\setlength{\captionheadmsw}{-10ex}
%\setcounter{page}{33}

%                    *            *           * 
%
%				Figure 1
%
\begin{figure}[h]

%\vspace*{\headroom}
\postscript{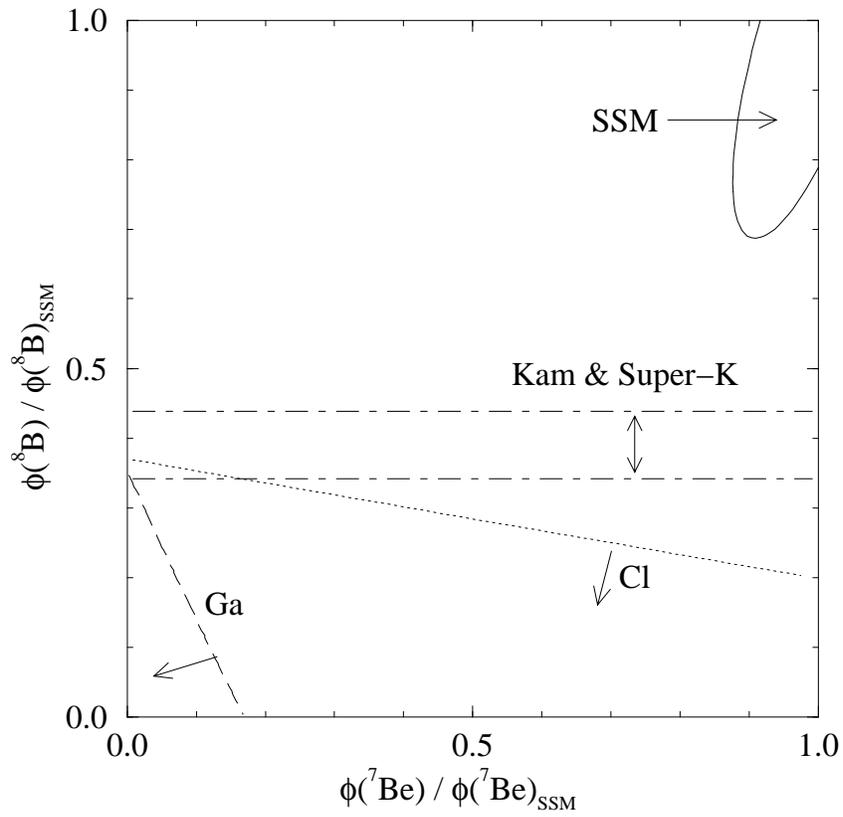}{1.00\hsize}
%\vspace{\psfigskip}
\vspace*{\captionhead}

\caption{
The constraints on the $^7$Be and $^8$B fluxes at 90\% C.L. from the
Homestake result (below the dotted line), the combined Kamiokande and
Super-Kamiokande results (between the dot-dashed lines), and the
combined SAGE and GALLEX results (below the dashed line).  The SSM
range is also shown (solid line, 90\% C.L.)}
\label{Fig:BeB_each}
\end{figure}

\clearpage

%                    *            *           * 
%
%				Figure 2
%
\begin{figure}[h]

%\vspace*{\headroom}
\postscript{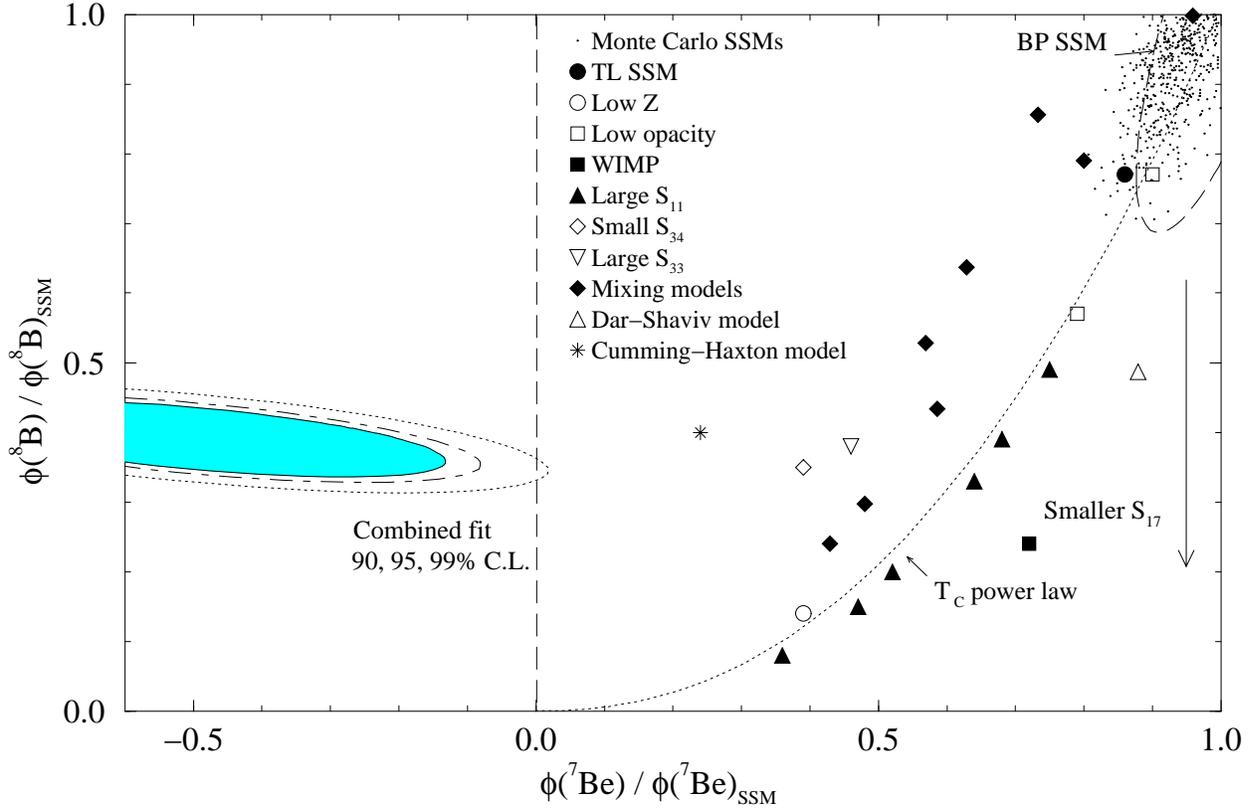}{1.00\hsize}
%\vspace{\psfigskip}
\vspace*{\captionhead}

\caption{
The constraints from the combined Cl, Ga, and \v{C}erenkov experiments
at 90, 95, and 99\% C.L.\   Also shown are the Bahcall-Pinsonneault SSM
region at 90\% C.L. \protect\cite{Bahcall-Pinsonneault-95}, the core
temperature power law and standard and nonstandard solar models
including the recent $^3$He diffusion model by Cunning and Haxton
\protect\cite{Cumming-Haxton} (see
Ref.~\protect\cite{Hata-Langacker-95} for references for the other
models).  A smaller $S_{17}$ cross section moves the solar model
predictions to a smaller $^8$B flux as indicted by the arrow.}
\label{Fig:BeB_comb_cnozero}
\end{figure}

\clearpage

%                    *            *           * 
%
%				Figure 3
%
\begin{figure}[h]

%\vspace*{\headroom}
\postscript{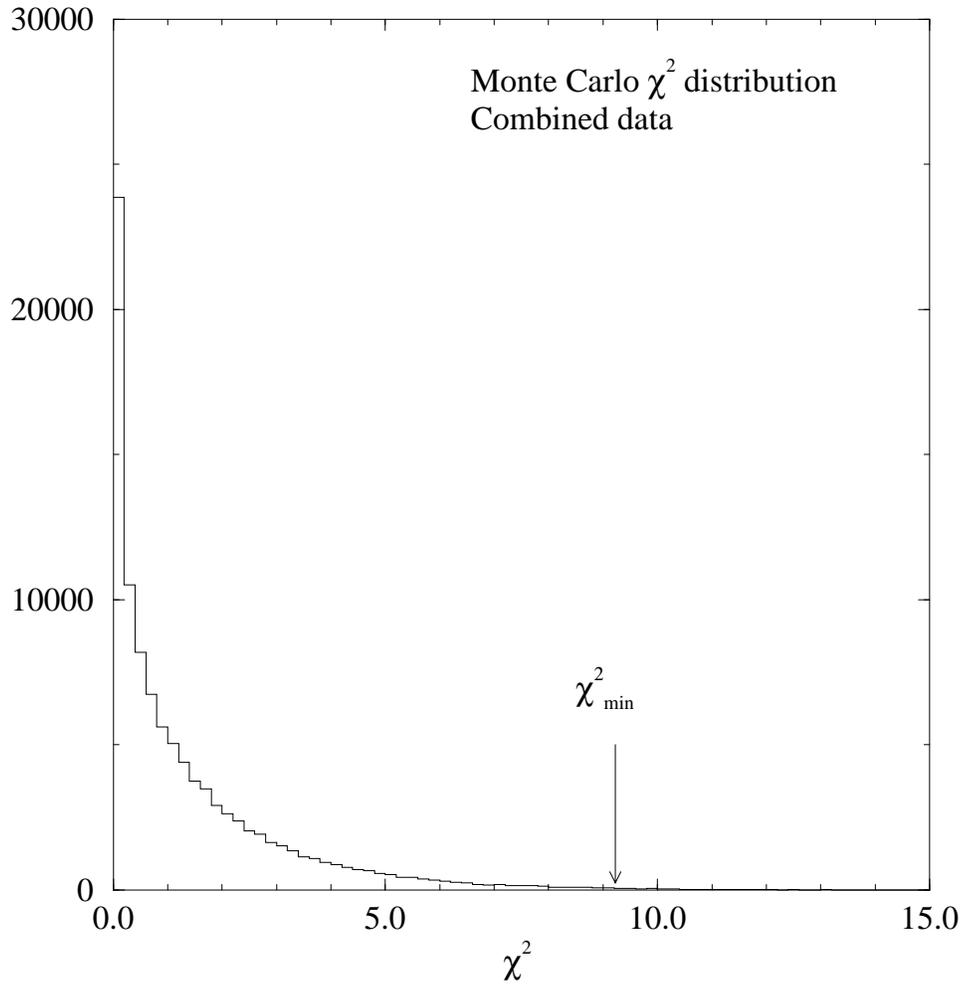}{1.00\hsize}
%\vspace{\psfigskip}
\vspace*{\captionhead}

\caption{
The Monte-Carlo distribution of $\chi^2$ minima when the best fit
fluxes in the physical region ($\phi({\rm Be})/\phi({\rm Be})_{\rm
SSM} = 0$ and $\phi({\rm B})/\phi({\rm B})_{\rm SSM} = 0.35$) are
assumed.  The actual $\chi^2$ minimum of the combined observations are
indicated by the arrow.  The best fit astrophysical solution is
excluded at the 99.4\% C.L.}
\label{Fig:gof_lin_comb}
\end{figure}

\clearpage

%                    *            *           * 
%
%				Figure 4
%
\begin{figure}[h]

%\vspace*{\headroom}
\postscript{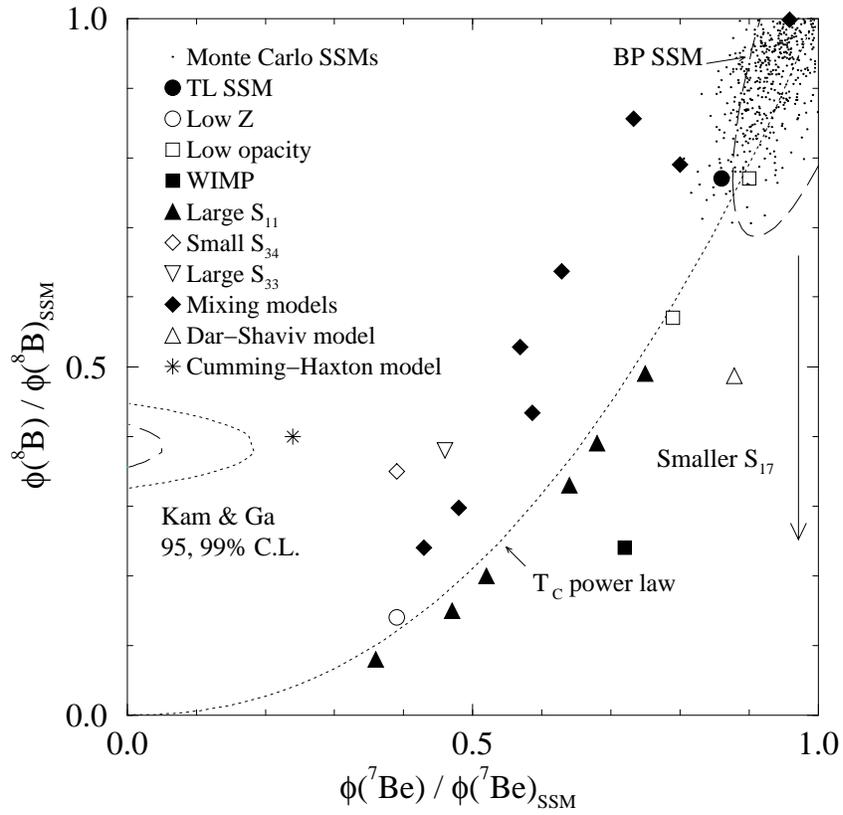}{1.00\hsize}
%\vspace{\psfigskip}
\vspace*{\captionhead}

\caption{
The result with the water \v{C}erenkov and gallium data only.}
\label{Fig:BeB_kam+ga}
\end{figure}

\clearpage

%                    *            *           * 
%
%				Figure 5
%
\begin{figure}[h]

%\vspace*{\headroom}
\postscript{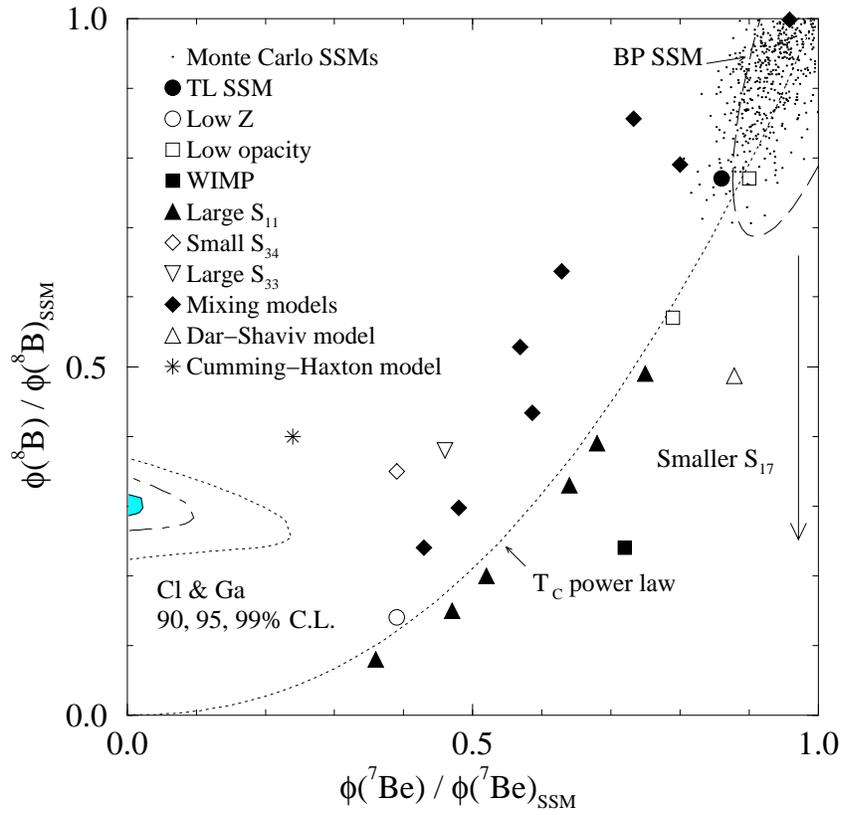}{1.00\hsize}
%\vspace{\psfigskip}
\vspace*{\captionhead}

\caption{
The result with the Homestake and gallium data only.}
\label{Fig:BeB_cl+ga}
\end{figure}

\clearpage

%                    *            *           * 
%
%				Figure 6
%
\begin{figure}[h]

%\vspace*{\headroom}
\postscript{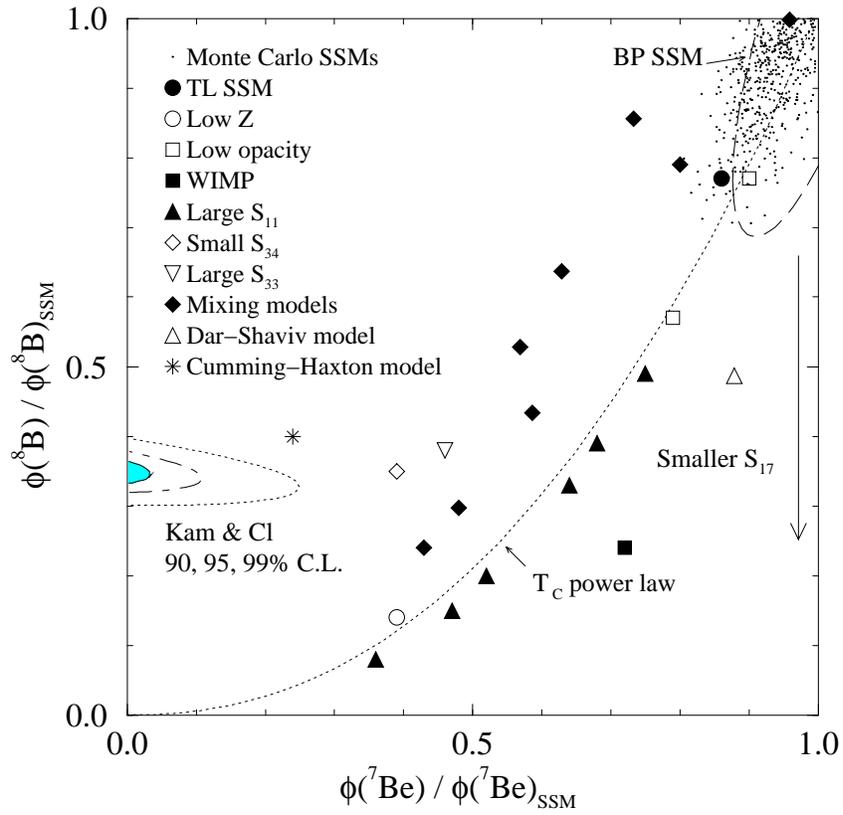}{1.00\hsize}
%\vspace{\psfigskip}
\vspace*{\captionhead}

\caption{
The result with the Homestake and water \v{C}erenkov data only.}
\label{Fig:BeB_kam+cl}
\end{figure}

\clearpage

%                    *            *           * 
%
%				Figure 7
%
\begin{figure}[h]

%\vspace*{\headroom}
\postscript{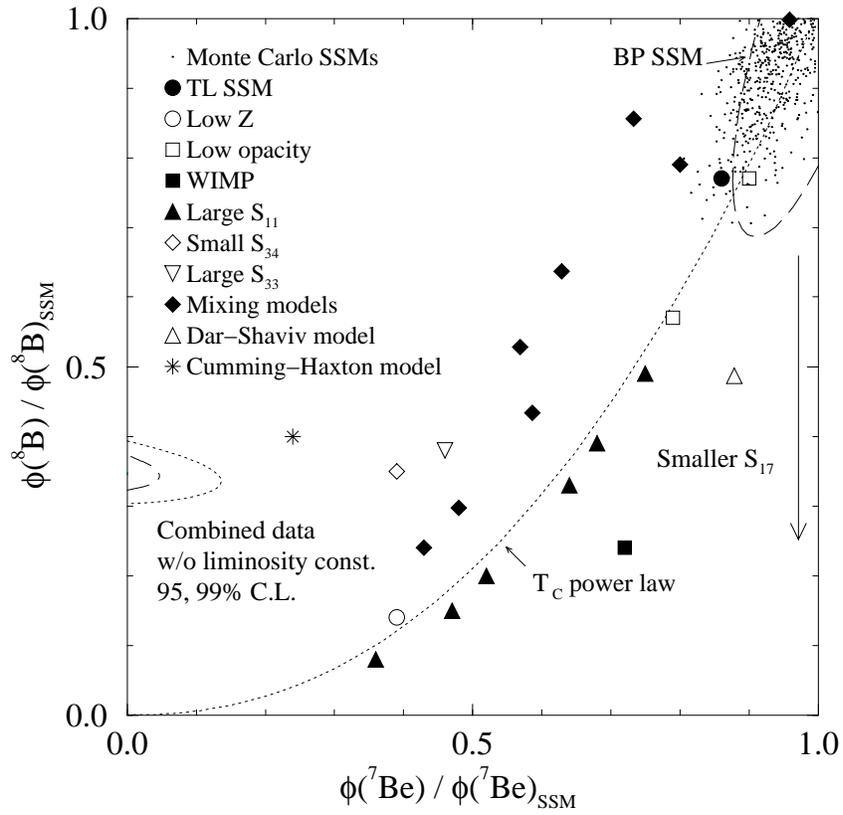}{1.00\hsize}
%\vspace{\psfigskip}
\vspace*{\captionhead}

\caption{
The result with the combined data but without imposing the luminosity}
constraint.
\label{Fig:BeB_comb_nolumi}
\end{figure}

\clearpage

%                    *            *           * 
%
%				Figure 8
%
\begin{figure}[h]

%\vspace*{\headroom}
\postscript{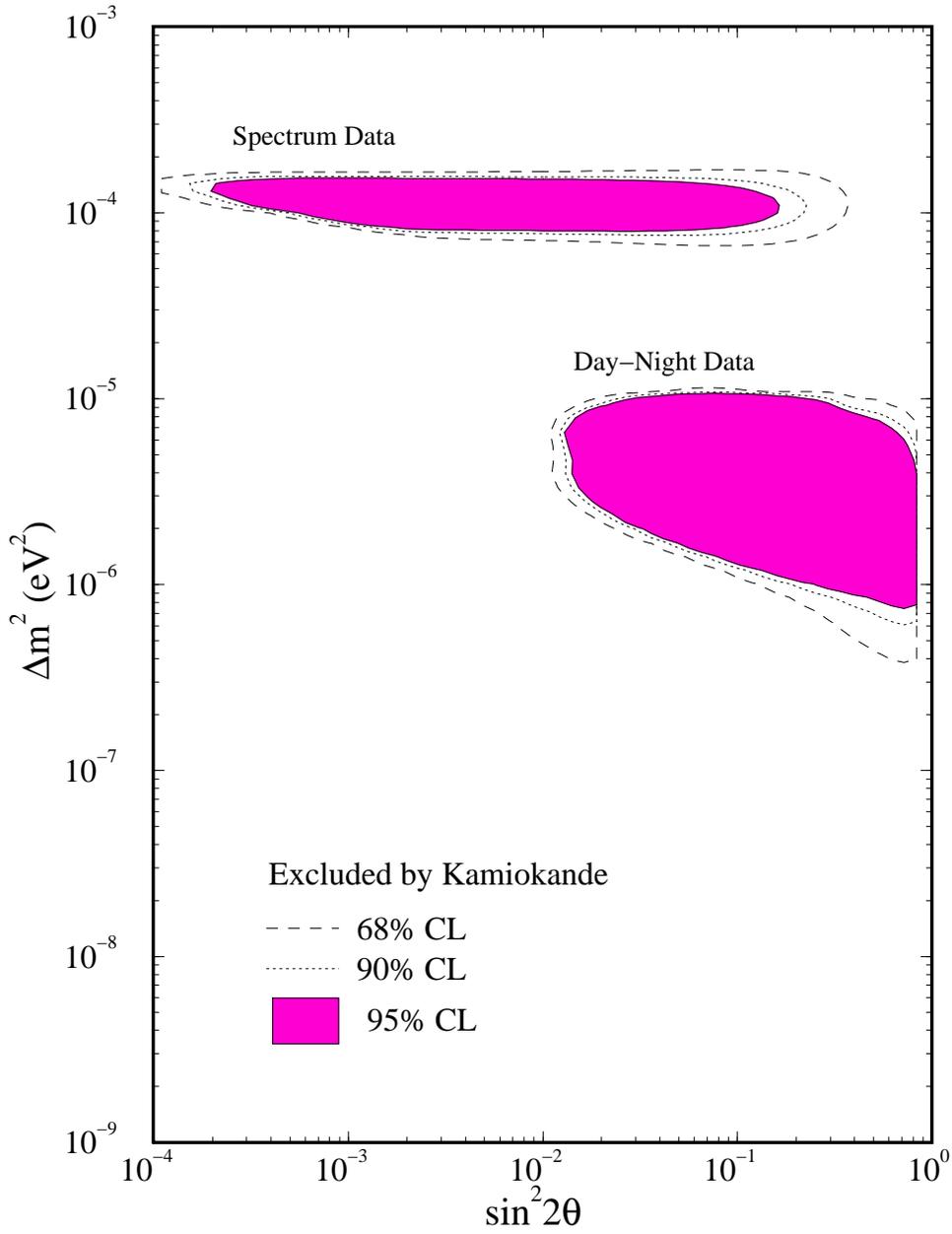}{1.00\hsize}
%\vspace{\psfigskip}
\vspace*{\captionheadmsw}

\caption{
The MSW parameter space excluded by the Kamiokande spectrum data
and day-night data.}
\label{Fig:k_exclusion}
\end{figure}

\clearpage

%                    *            *           * 
%
%				Figure 9
%
\begin{figure}[h]

%\vspace*{\headroom}
\postscript{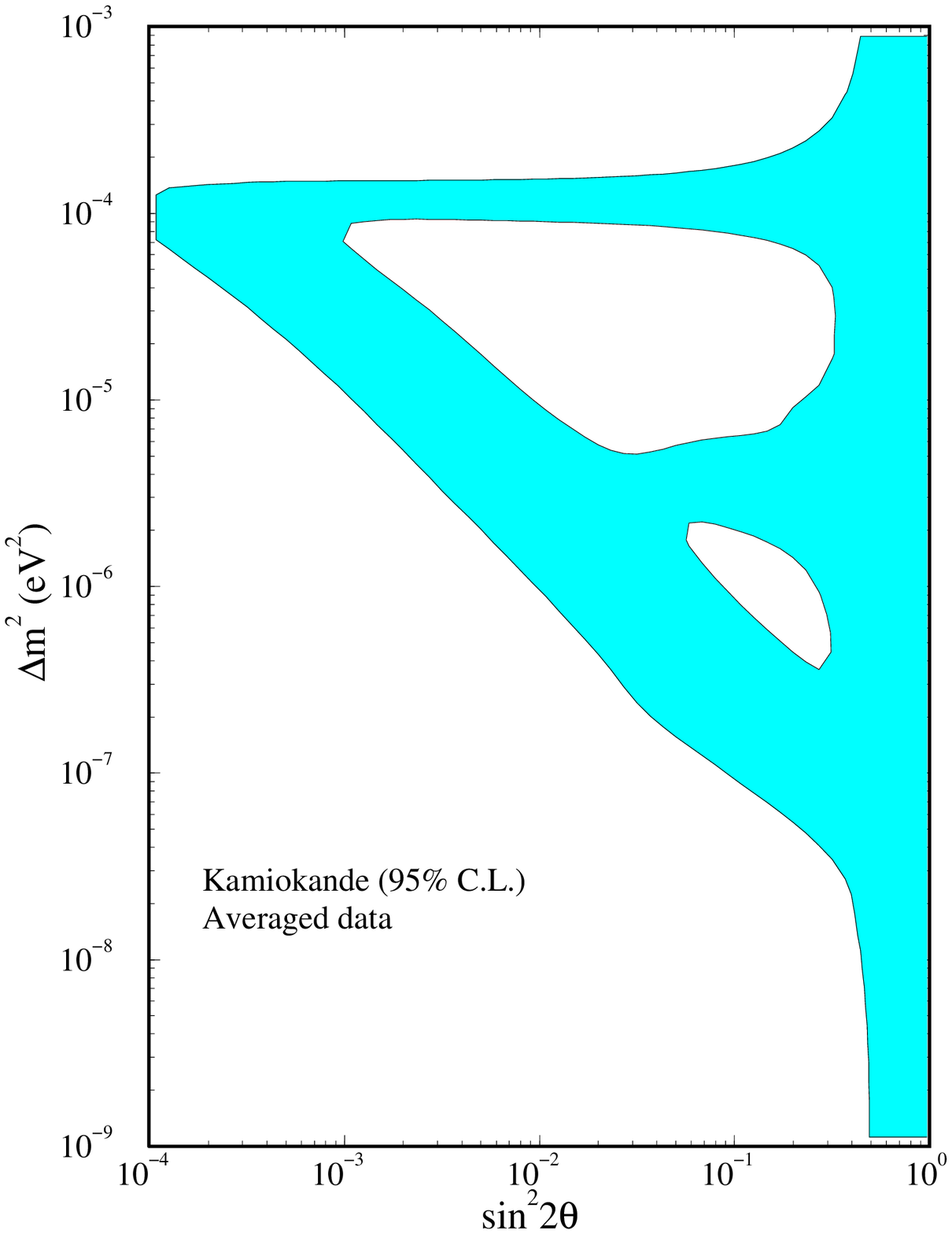}{1.00\hsize}
%\vspace{\psfigskip}
\vspace*{\captionheadmsw}

\caption{
The MSW parameter space allowed by the Kamiokande total rate.}
\label{Fig:fkam_ave}
\end{figure}

\clearpage

%                    *            *           * 
%
%				Figure 10
%
\begin{figure}[h]

%\vspace*{\headroom}
\postscript{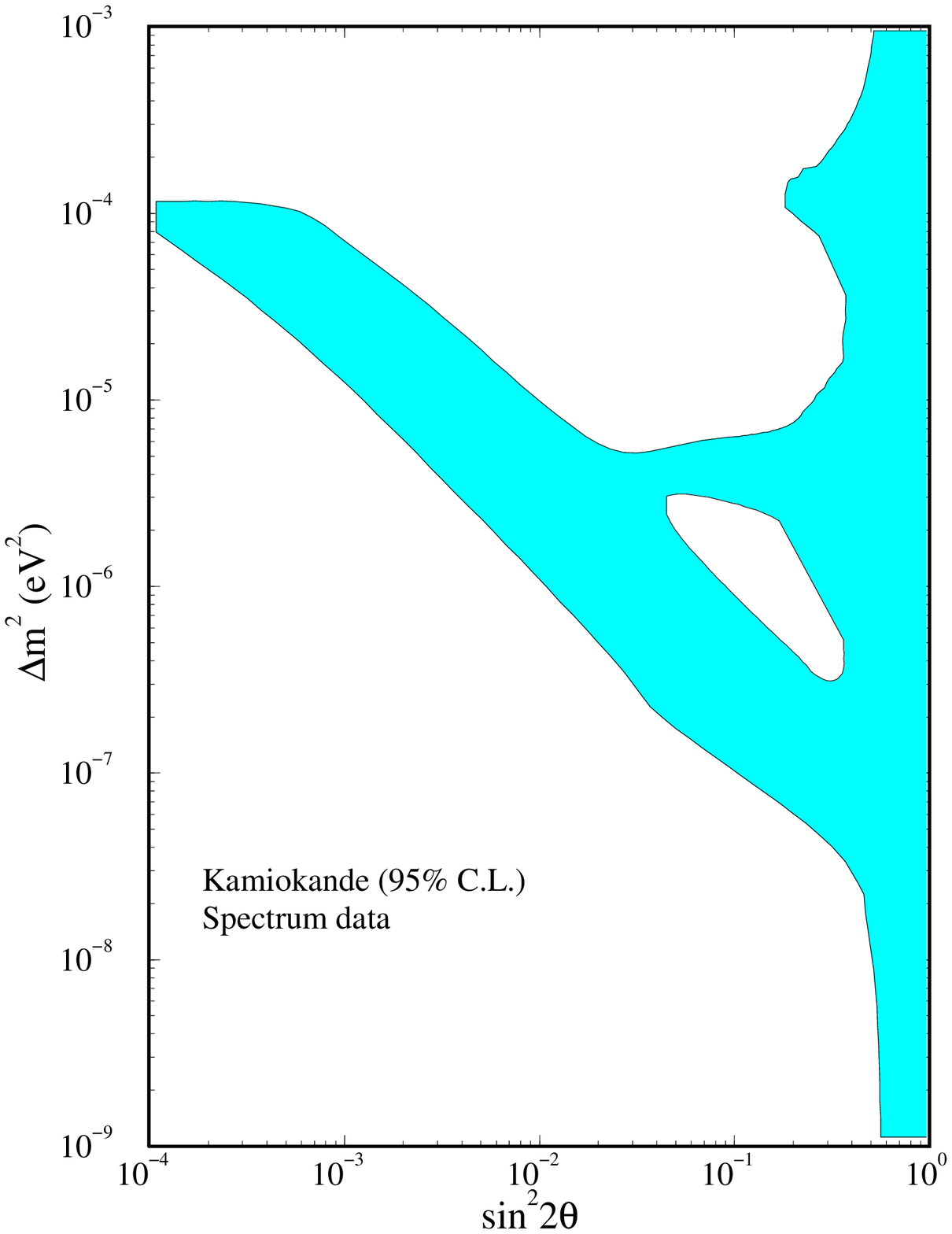}{1.00\hsize}
%\vspace{\psfigskip}
\vspace*{\captionheadmsw}

\caption{
The MSW parameter space allowed by the Kamiokande total rate and
spectrum data.}
\label{Fig:fkam_sp}
\end{figure}

\clearpage

%                    *            *           * 
%
%				Figure 11
%
\begin{figure}[h]

%\vspace*{\headroom}
\postscript{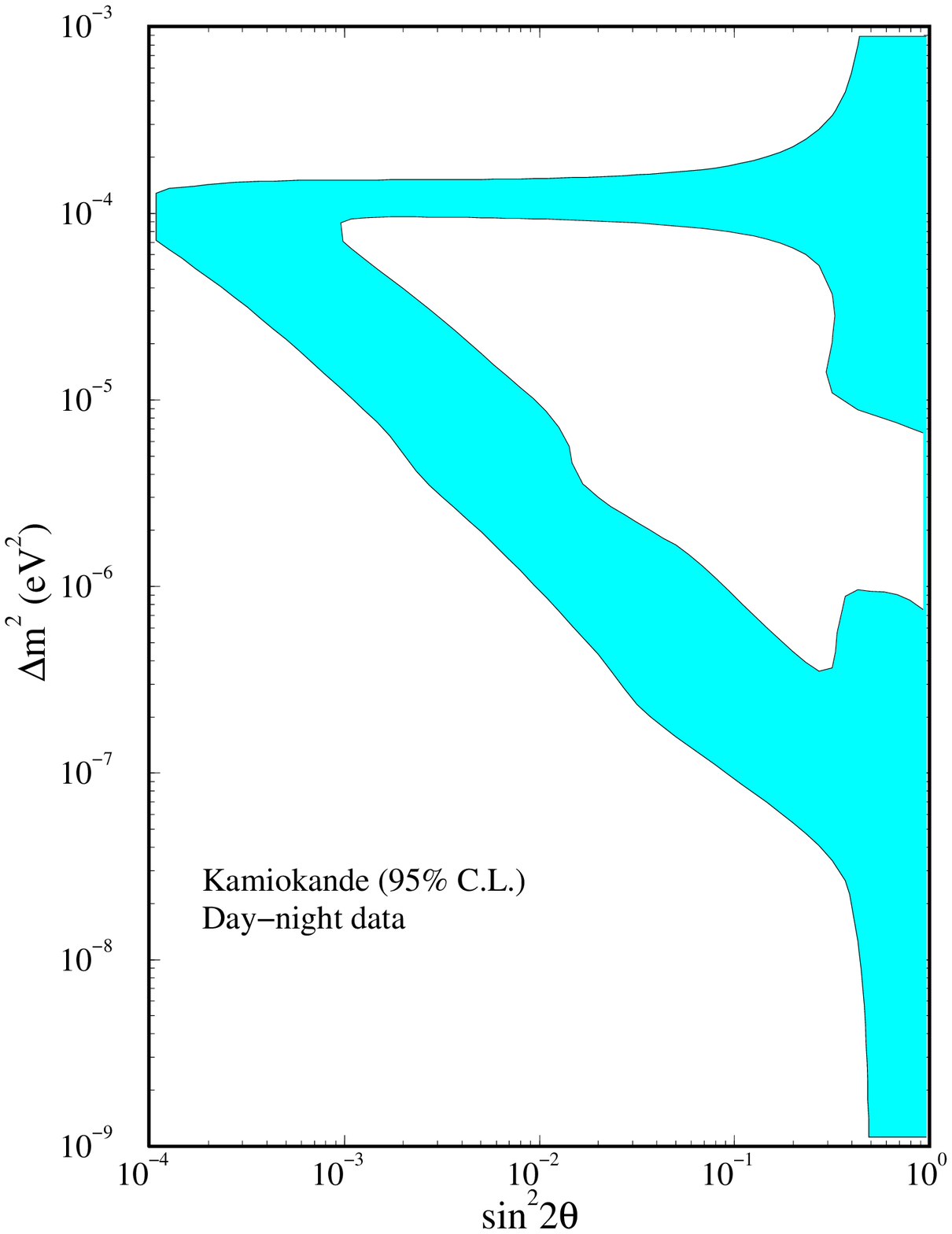}{1.00\hsize}
%\vspace{\psfigskip}
\vspace*{\captionheadmsw}

\caption{
The MSW parameter space allowed by the Kamiokande total rate and
day-night data.}
\label{Fig:fkam_dn}
\end{figure}

\clearpage

%                    *            *           * 
%
%				Figure 12
%
\begin{figure}[h]

%\vspace*{\headroom}
\postscript{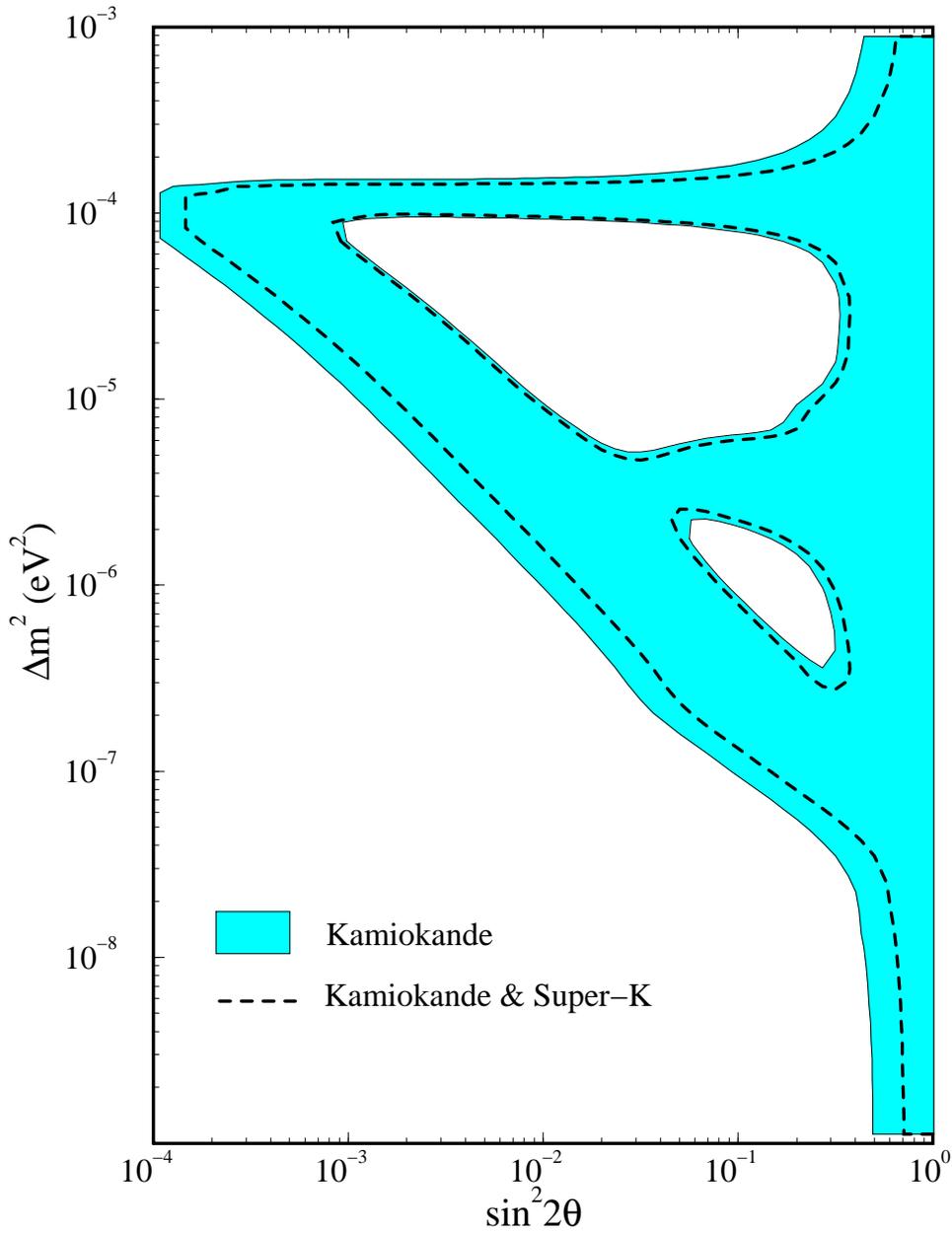}{1.00\hsize}
%\vspace{\psfigskip}
\vspace*{\captionheadmsw}

\caption{
The comparison of the MSW parameter space allowed by the Kamiokande
total rate (shaded region) and the combined Kamiokande and preliminary
Super-Kamiokande rate (dashed lines).  The theory error ($\sim$ 15\%)
is the leading uncertainty in the combined fits.}
\label{Fig:fkam_comb}
\end{figure}

\clearpage

%                    *            *           * 
%
%				Figure 13
%
\begin{figure}[h]

%\vspace*{\headroom}
\postscript{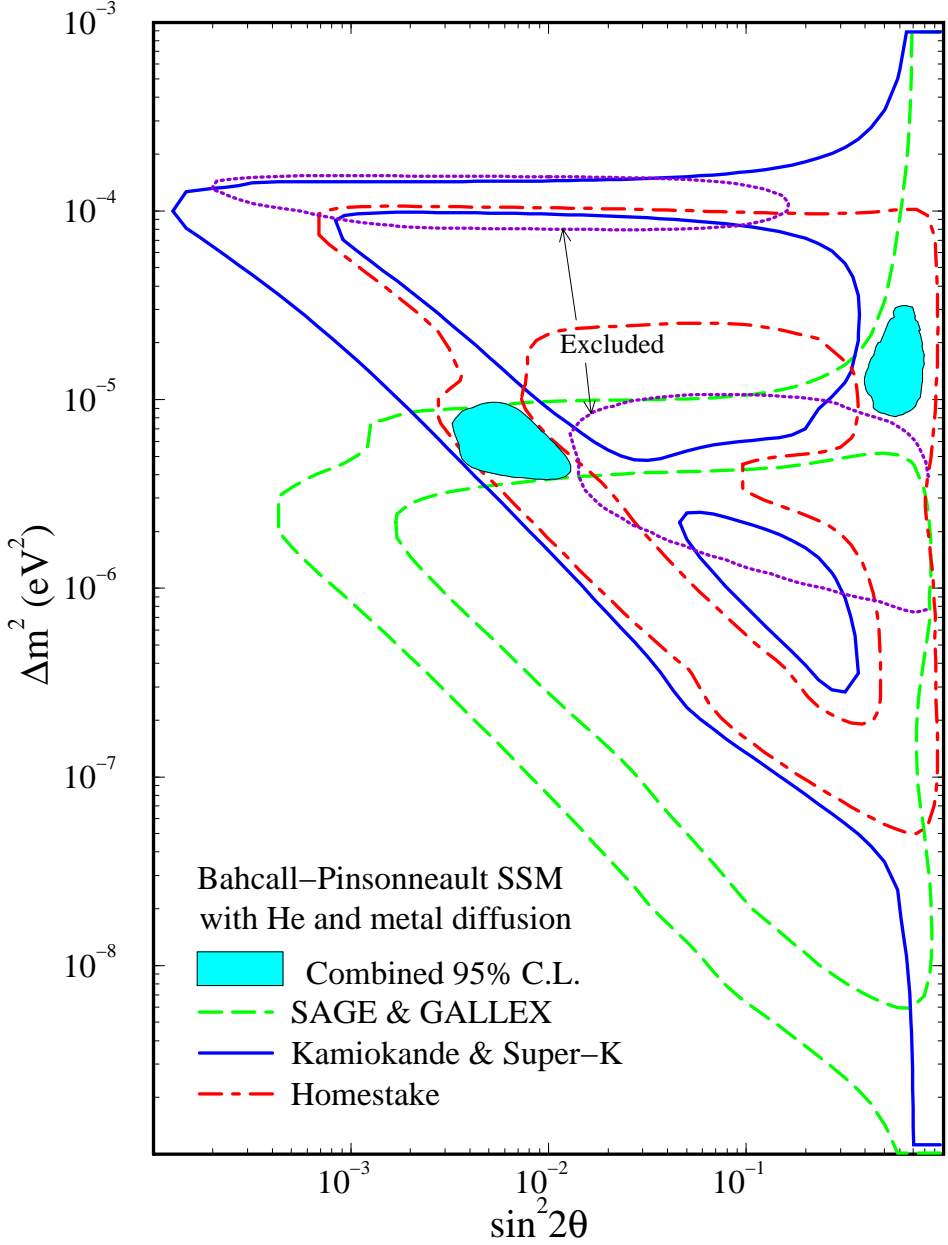}{1.00\hsize}
%\vspace{\psfigskip}
\vspace*{\captionheadmsw}

\caption{
The result of the MSW parameter space (shaded regions) allowed by the
combined observations at 95\% C.L.\ assuming the Bahcall-Pinsonneault
SSM with He diffusion.  The constraints from Homestake, combined
Kamiokande and Super-Kamiokande, and combined SAGE and GALLEX are
shown by the dot-dashed, solid, and dashed lines, respectively.  Also
shown are the regions excluded by the Kamiokande spectrum and
day-night data (dotted lines).}
\label{Fig:p_comb}
\end{figure}

\clearpage

%                    *            *           * 
%
%				Figure 14
%
\begin{figure}[h]

%\vspace*{\headroom}
\postscript{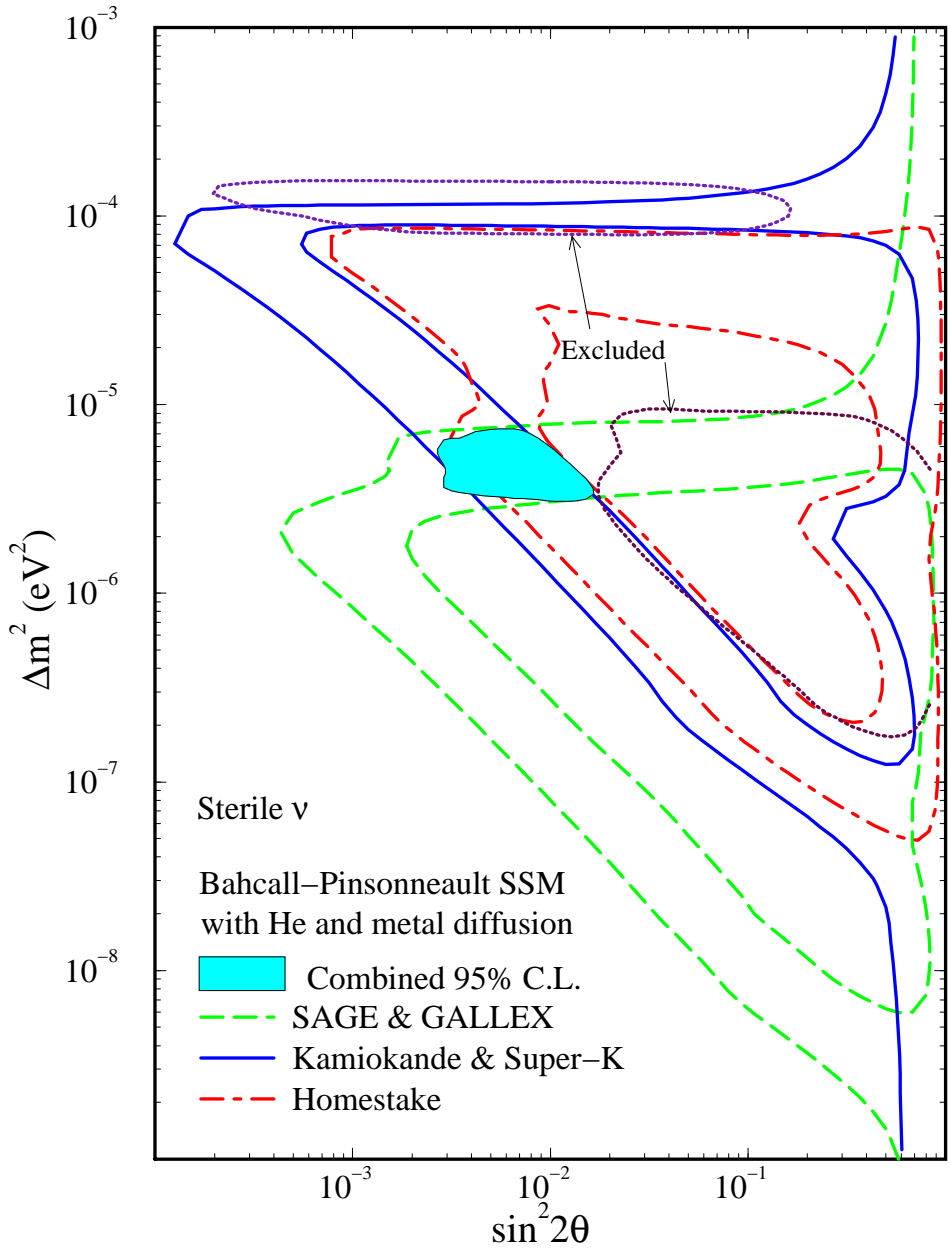}{1.00\hsize}
%\vspace{\psfigskip}
\vspace*{\captionheadmsw}

\caption{
Same as Fig.~\ref{Fig:p_comb} except that this is for oscillations
to sterile neutrinos.  }
\label{Fig:p_comb_sterile}
\end{figure}

\clearpage

%                    *            *           * 
%
%				Figure 15
%
\begin{figure}[h]

%\vspace*{\headroom}
\postscript{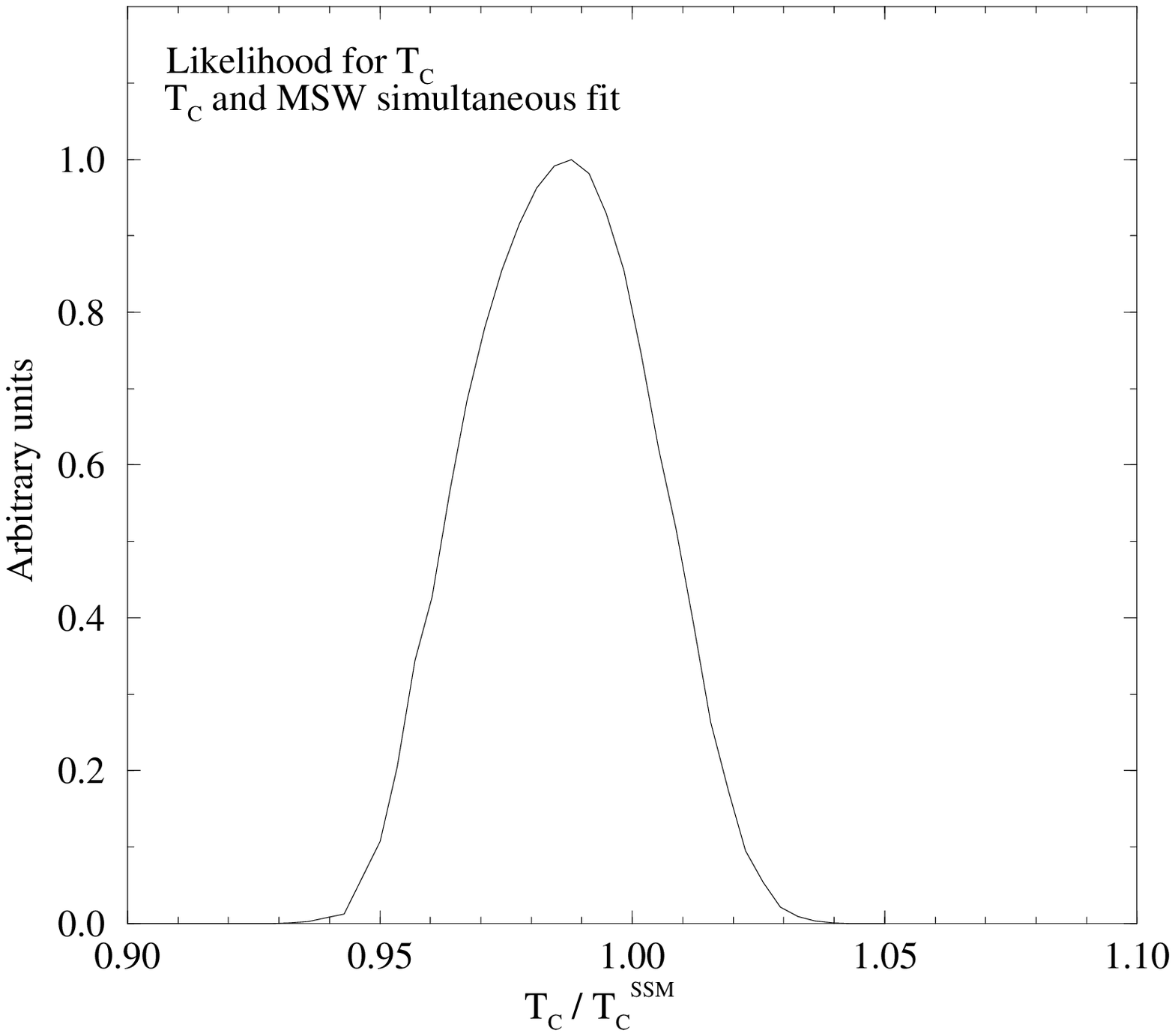}{1.00\hsize}
%\vspace{\psfigskip}
\vspace*{\captionhead}

\caption{
The likelihood distribution of the core temperature from the
simultaneous MSW fit to the combined observations.}
\label{Fig:tc-P}
\end{figure}

\clearpage

%                    *            *           * 
%
%				Figure 16
%
\begin{figure}[h]

%\vspace*{\headroom}
\postscript{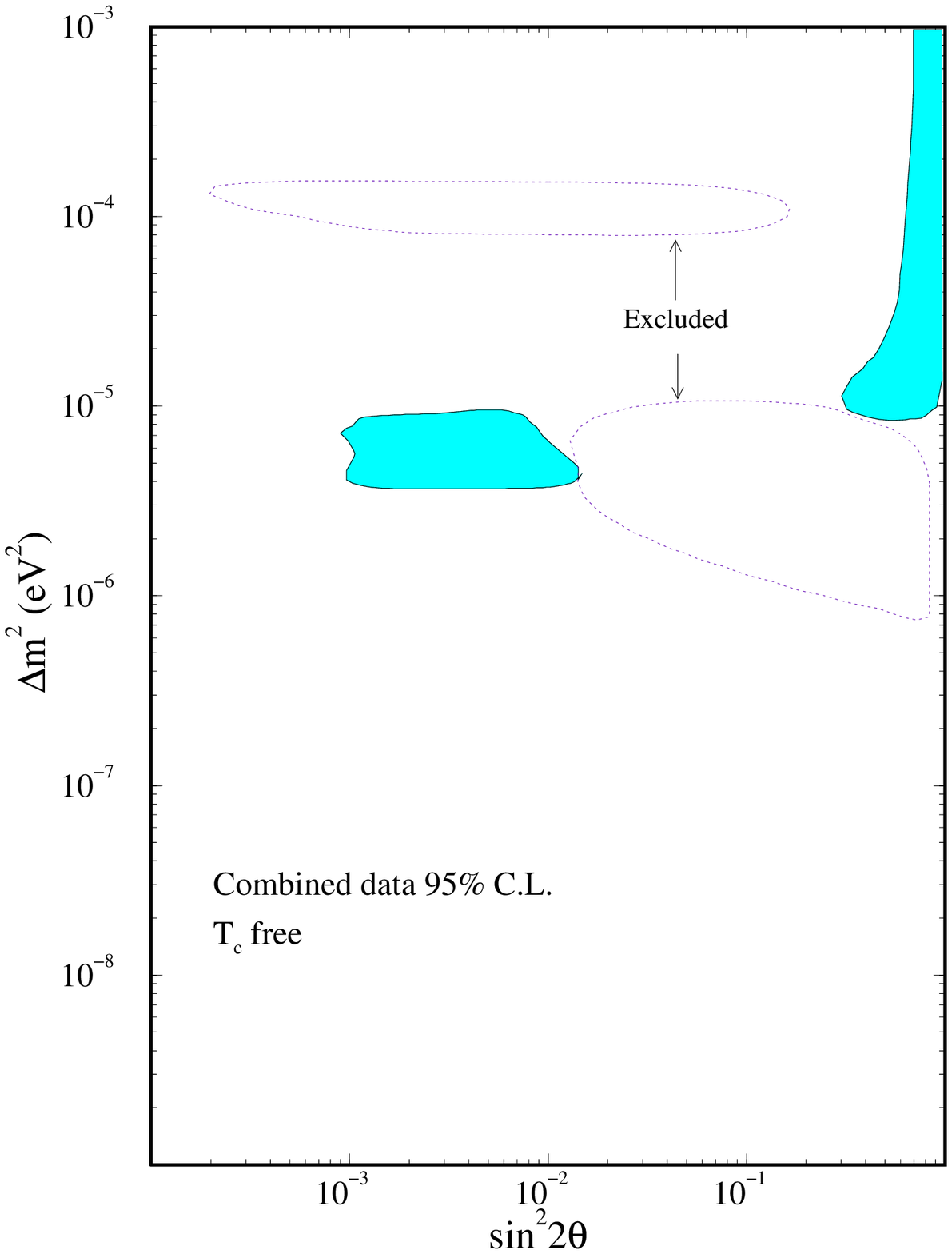}{1.00\hsize}
%\vspace{\psfigskip}
\vspace*{\captionheadmsw}

\caption{
The MSW parameter space allowed by the combined observations when
the core temperature is used as a free parameter.  The model
independent exclusion regions by the Kamiokande spectrum and day-night
data are also shown.  }
\label{Fig:p_comb_tcfree}
\end{figure}

\clearpage

%                    *            *           * 
%
%				Figure 17
%
\begin{figure}[h]

%\vspace*{\headroom}
\postscript{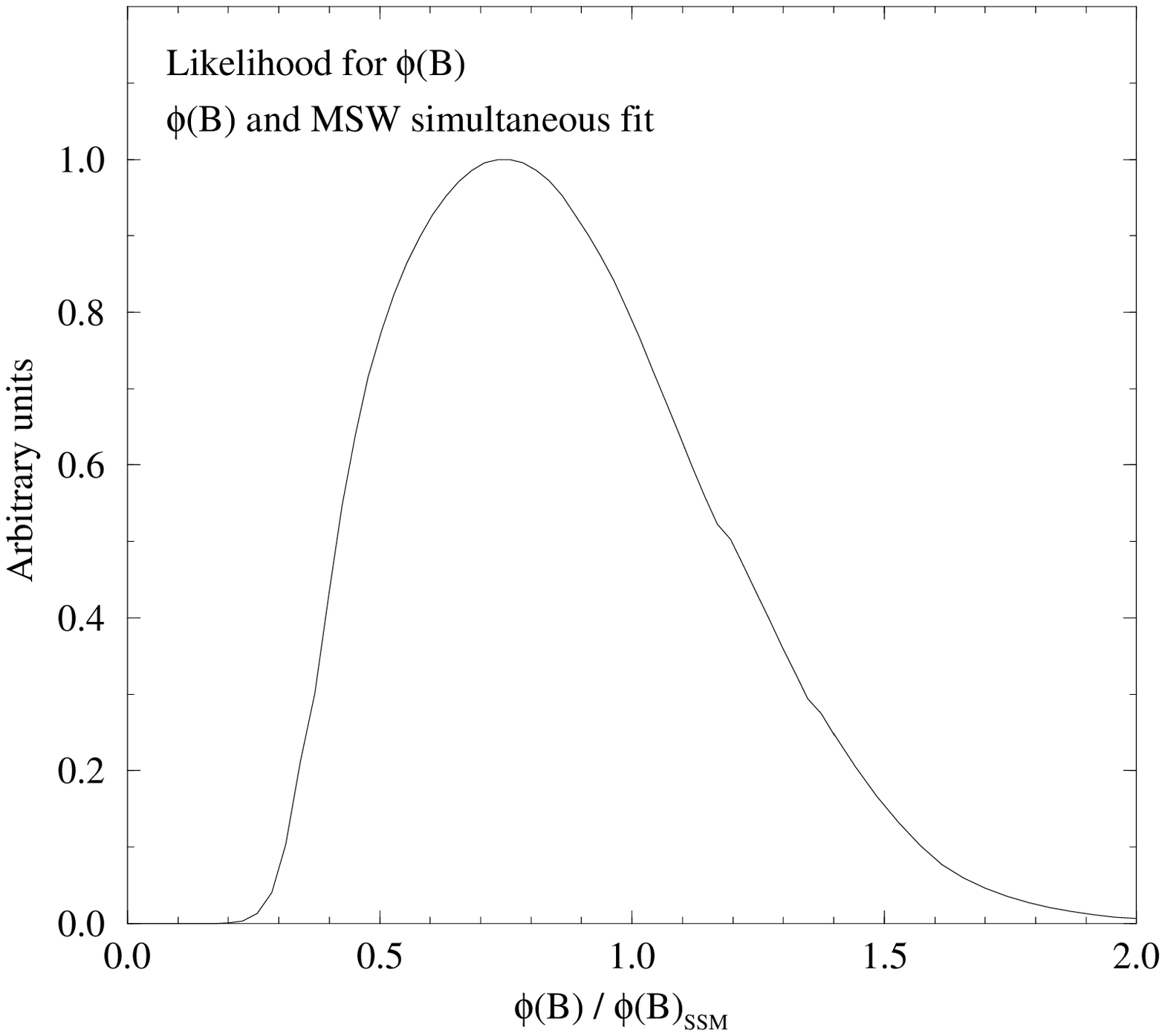}{1.00\hsize}
%\vspace{\psfigskip}
\vspace*{\captionhead}

\caption{
The likelihood distribution of the core temperature from the
simultaneous MSW fit to the combined observations.}
\label{Fig:B-P}
\end{figure}

\clearpage

%                    *            *           * 
%
%				Figure 18
%
\begin{figure}[h]

%\vspace*{\headroom}
\postscript{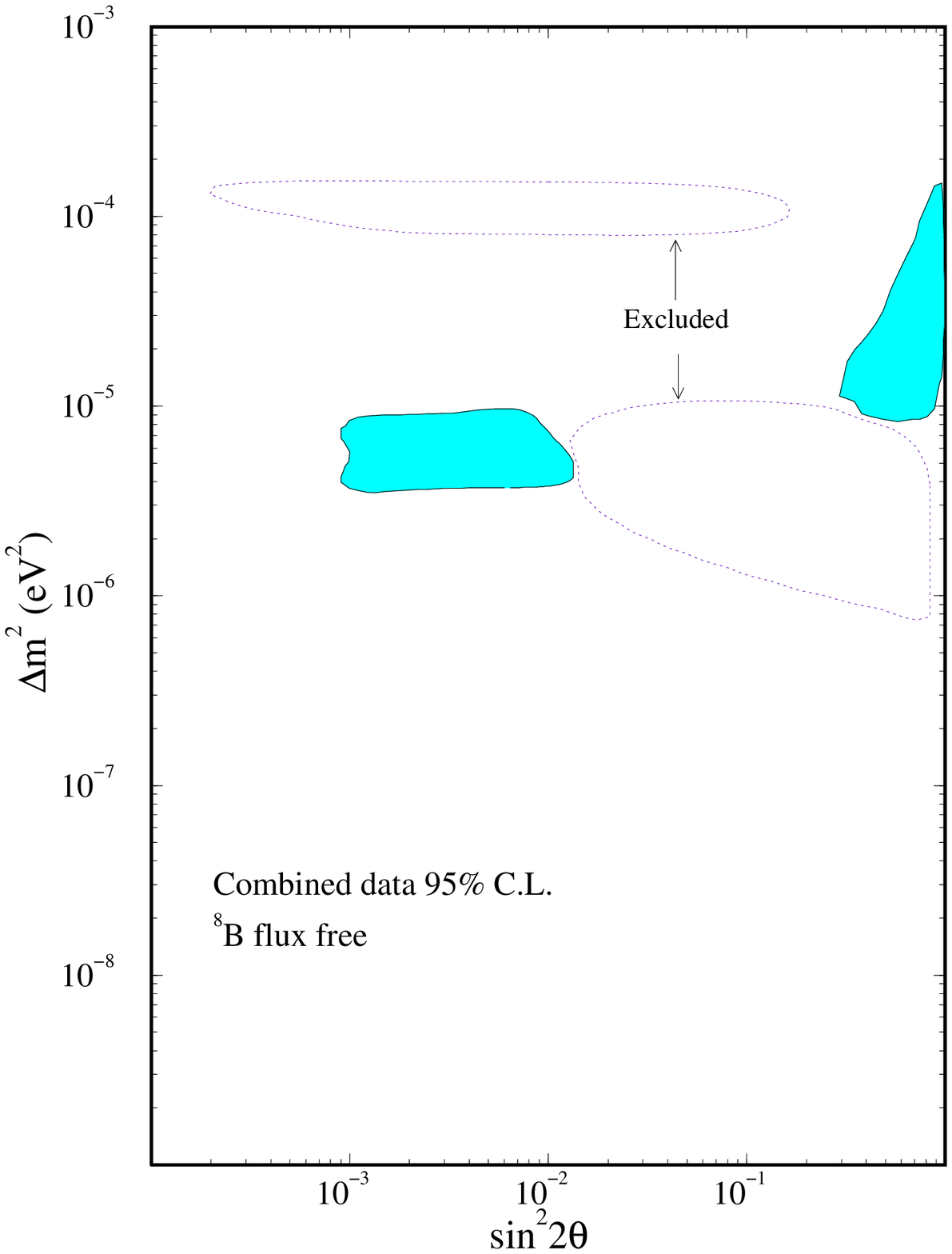}{1.00\hsize}
%\vspace{\psfigskip}
\vspace*{\captionheadmsw}

\caption{
The MSW parameter space allowed by the combined observations when the
$^8$B flux is used as a free parameter.  The model independent
exclusion regions by the Kamiokande spectrum and day-night data are
also shown.}
\label{Fig:p_comb_Bfree}
\end{figure}

\clearpage

%                    *            *           * 
%
%				Figure 19
%
\begin{figure}[h]

%\vspace*{\headroom}
\postscript{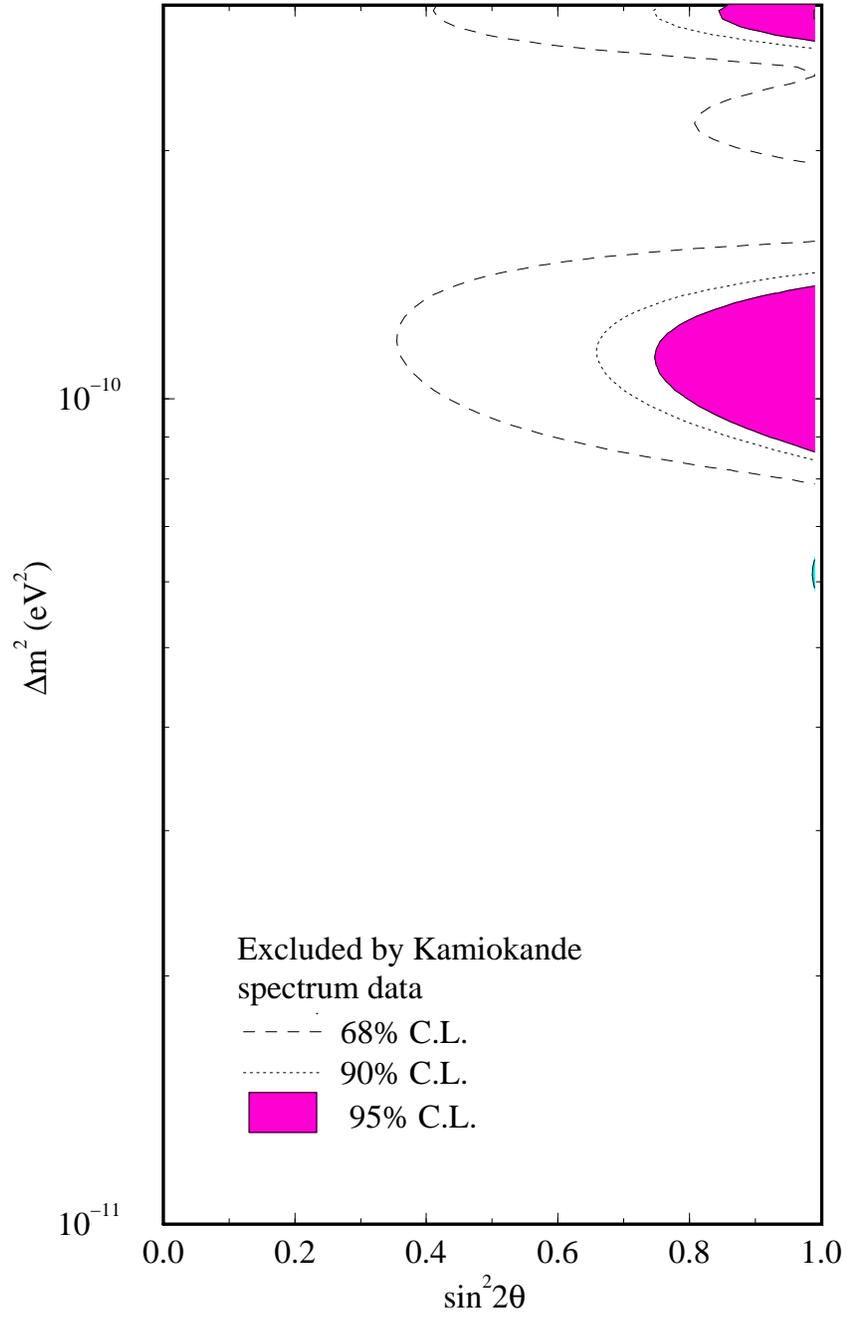}{1.00\hsize}
%\vspace{\psfigskip}
\vspace*{\captionheadmsw}

\caption{
The vacuum oscillation parameter space excluded by the Kamiokande
spectrum data.  }
\label{Fig:vexclusion_kam_sp}
\end{figure}

\clearpage

%                    *            *           * 
%
%				Figure 20
%
\begin{figure}[h]

%\vspace*{\headroom}
\postscript{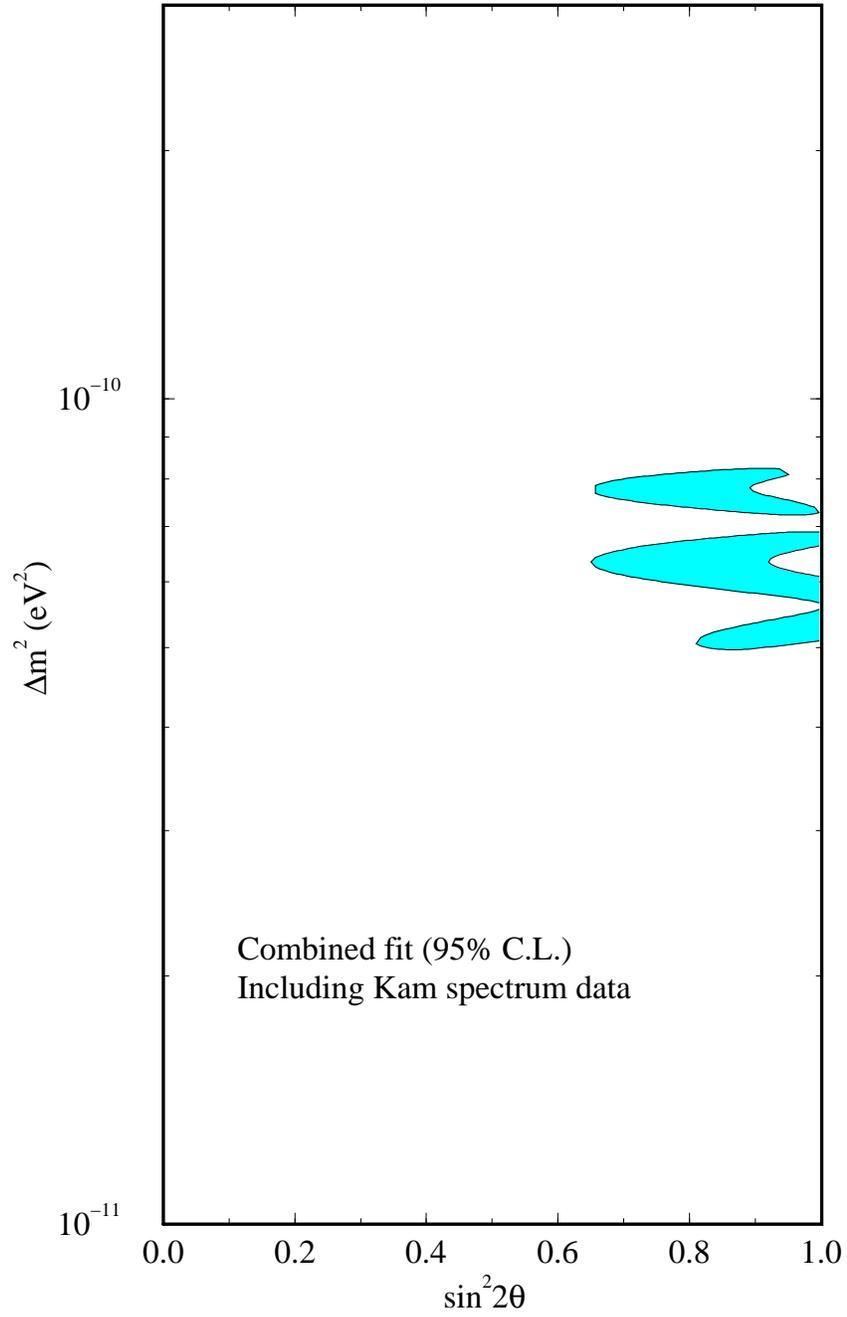}{1.00\hsize}
%\vspace{\psfigskip}
\vspace*{\captionheadmsw}

\caption{
The vacuum oscillation parameter space allowed by the combined
observations including the Kamiokande spectrum data.}
\label{Fig:vfcomb+sk+kam_sp}
\end{figure}

\clearpage

%                    *            *           * 
%
%				Figure 21
%
\begin{figure}[h]

%\vspace*{\headroom}
\postscript{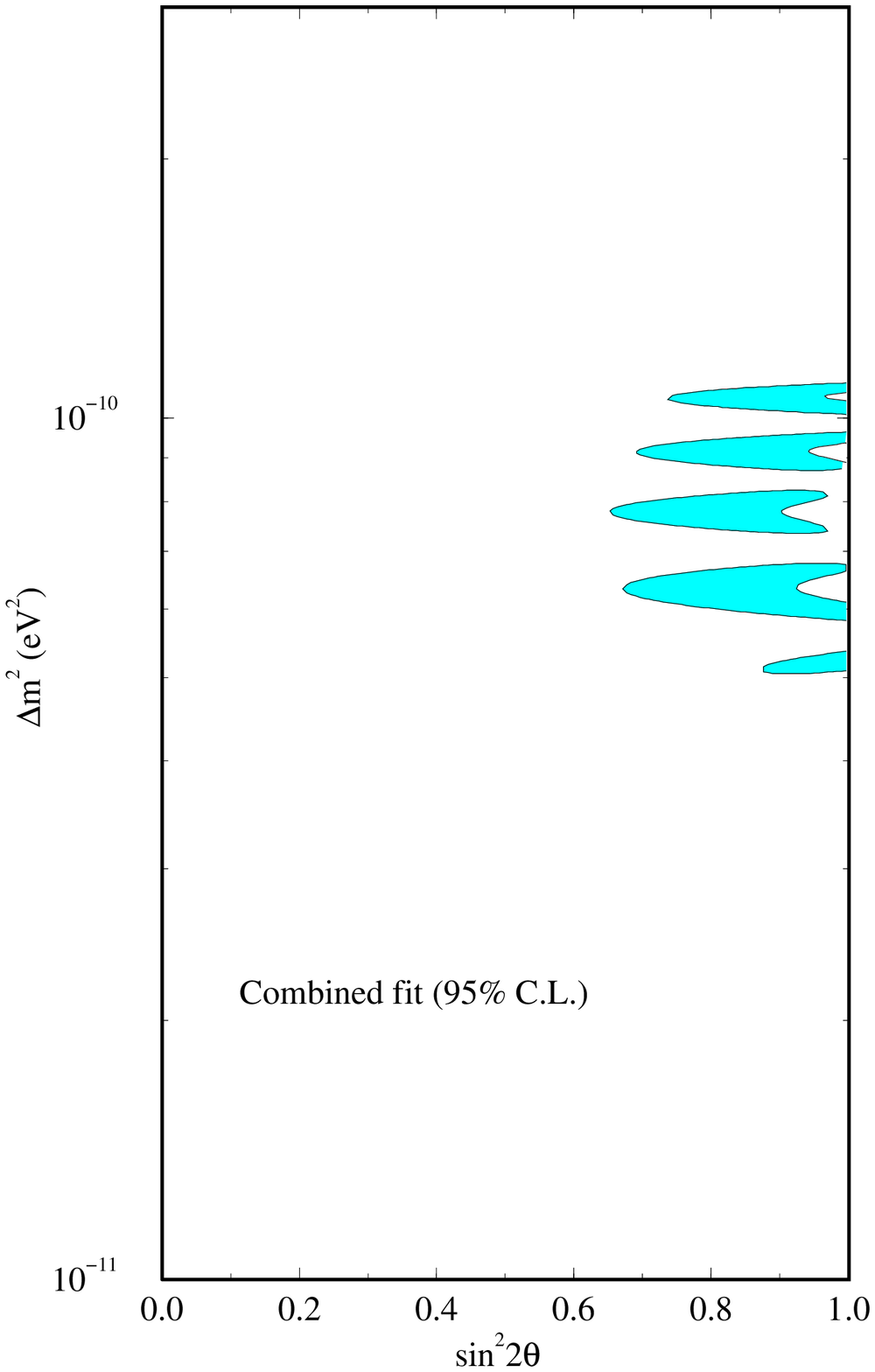}{1.00\hsize}
%\vspace{\psfigskip}
\vspace*{\captionheadmsw}

\caption{
The vacuum oscillation parameter space allowed by the combined
observations but without the Kamiokande spectrum data.}
\label{Fig:vfcomb+sk}
\end{figure}

\clearpage


\begin{thebibliography}{99}


\bibitem{MSW}
L.\ Wolfenstein, 
Pays.\ Rev.\ D {\bf 17}, 2369 (1978); {\bf 20}, 2634 (1979);
%
S.\ P.\ Mikheyev and A.\ Yu.\ Smirnov, 
Yad.\ Fiz.\ {\bf 42}, 1441 (1985) 
[Sov.\ J.\ Nucl.\ Phys. {\bf 42}, 913 (1985)]; 
Nuovo Cimento {\bf 9C}, 17 (1986).

\bibitem{Homestake}
B.\ T.\ Cleveland {\it et al.}, Preprint, 1996.

\bibitem{Kamiokande}
Kamiokande II Collaboration, K.\ S.\ Hirata {\it et al.}, 
Phys.\ Rev.\ Lett.\ {\bf 65}, 1297 (1990); 
{\bf 65}, 1301(1990); {\bf 66}, 9 (1991);
Phys.\ Rev.\ D {\bf 44}, 2241 (1991);
Kamiokande III collaboration, Y.\ Fukuda {\it et al.}, 
Phys.\ Rev.\ Lett.\ {\bf 77}, 1683 (1996). 

\bibitem{Super-Kamiokande}
Y.\ Totsuka, to be published in {\it Proceedings for Texas Conference},
December 1996.

\bibitem{SAGE} 
SAGE Collaboration, A.\ I.\ Abazov, {\it et al.}, 
Phys.\ Rev.\ Lett.\ {\bf 67}, 3332 (1991);
J.\ N.\ Abdurashitov {\it et al.}, 
Phys.\ Lett.\ B {\bf 328}, 234 (1994);
Phys.\ Rev.\ Lett.\ {\bf 77}, 4708 (1996).

\bibitem{GALLEX} 
GALLEX Collaboration, P.\ Anselmann {\it et al.},
Phys.\ Lett.\ B {\bf 285}, 376 (1992); 
{\bf 285}, 390 (1992);
{\bf 314}, 445 (1993);
{\bf 327}, 337 (1994);
{\bf 357}, 237 (1995);
W.\ Hampel {\it et al.},
Phys.\ Lett.\ B {\bf 388}, 384 (1996).

\bibitem{Bahcall-Pinsonneault-95}
J.\ N.\ Bahcall and M.\ H.\ Pinsonneault, 
Rev.\ Mod.\ Phys.\ {\bf 67}, 781 (1995).

\bibitem{Bahcall-Pinsonneault-Basu-Christensen-Dalsgaard}
J.\ N.\ Bahcall, M.\ H.\ Pinsonneault, S.\ Basu, and 
J.\ Christensen-Dalsgaard,
Phys.\ Rev.\ Lett.\ {\bf 78}, 171 (1996).

\bibitem{Bahcall-Kamionkowsky-Sirlin}
J.\ N.\ Bahcall, M.\ Kamionkowsky, and A.\ Sirlin,
Phys.\ Rev.\ D {\bf 51}, 6146 (1995). 

\bibitem{Bahcall-etal-96}
J.\ N.\ Bahcall, {\it et al.},
Phys.\ Rev.\ C {\bf 54}, 411 (1996).

\bibitem{Hata-Haxton}
N.\ Hata and W.\ Haxton,
Phys.\ Lett.\ B {\bf 353}, 422 (1995). 

\bibitem{BHKL}
S.\ Bludman, N.\ Hata, D.\ Kennedy, and P.\ Langacker, 
Phys.\ Rev.\ D {\bf 47}, 2220 (1993).

\bibitem{Hata-Bludman-Langacker}
N.\ Hata, S.\ Bludman, and P.\ Langacker,
Phys.\ Rev.\ D {\bf 49}, 3622 (1994).

\bibitem{Hata-Langacker-93}
N.\ Hata and P.\ Langacker,
Phys.\ Rev.\ D {\bf 48}, 2937 (1993).

\bibitem{Hata-Langacker-94}
N.\ Hata and P.\ Langacker,
Phys.\ Rev.\ D {\bf 50}, 632 (1994).

\bibitem{Hata-Langacker-95}
N.\ Hata and P.\ Langacker,
Phys.\ Rev.\ D {\bf 52}, 420 (1995).

\bibitem{Fogli-Lisi}
G.\ L.\ Fogli and E.\ Lisi 
Phys.\ Rev.\ D {\bf 54}, 3667 (1996). 

\bibitem{Smirnov-96}
A.\ Yu Smirnov,
in {\it Proceedings for the 28th
International Conference on High-energy Physics (ICHEP 96)}, 
Warsaw, Poland, 25-31 July 1996 
(Los Alamos e-Print Archive No. hep-ph/9611465). 

\bibitem{Bahcall-Ulrich}
J.\ N.\ Bahcall and R.\ N.\ Ulrich, 
Rev.\ Mod.\ Phys.\ {\bf 60}, 297 (1988).

\bibitem{Bahcall-book}
J.\ N.\ Bahcall, 
{\it Neutrino Astrophysics}, (Cambridge University Press, Cambridge, 
England, 1989).

\bibitem{Bahcall-Bethe}
J.\ N.\ Bahcall and H.\ A.\ Bethe,
Phys.\ Rev.\ D {\bf 47}, 1298 (1993); 
Phys.\ Rev.\ Lett.\ {\bf 65}, 2233 (1990);
H.\ A.\ Bethe and J.\ N.\ Bahcall, 
Phys.\ Rev.\ D {\bf 44}, 2962 (1991).

\bibitem{Bludman-Kennedy-Langacker}
S.\ Bludman, D.\ Kennedy, and P.\ Langacker, 
Phys.\ Rev.\ D {\bf 45}, 1810 (1992); 
Nucl.\ Phys.\ B {\bf 374}, 373 (1992).

\bibitem{Spiro-Vignaud}
M.\ Spiro and D.\ Vignaud,
Phys.\ Lett.\ B {\bf 242}, 279-284 (1990).

\bibitem{Castellani-Degl'Innocenti-Fiorentini-AA}
V.\ Castellani, S.\ Degl'Innocenti, and G.\ Fiorentini,
Astron.\ Astrophys.\ {\bf 271}, 601 (1993).

\bibitem{Castellani-Degl'Innocenti-Fiorentini-PRD}
V.\ Castellani, S.\ Degl'Innocenti, and G.\ Fiorentini,
Phys.\ Rev.\ D {\bf 50}, 4749 (1994). 

\bibitem{Parke}
S.\ Parke,
Phys.\ Rev.\ Lett.\ {\bf 74}, 839 (1995). 

\bibitem{Heeger-Robertson}
K.\ M.\ Heeger and R.\ G.\ H.\ Robertson,
Phys.\ Rev.\ Lett,\ {\bf 77}, 3720 (1996). 

\bibitem{Numerical-Recipes}
W.\ H.\ Press, S.\ A.\ Teukolsky, W.\ T.\ Vetterling, and B.\ P.\ Flannery,
{\it Numerical Recipes in Fortran, The Art of Scientific Computing, 2nd ed.}
(Cambridge University Press, Cambridge, 1992.) 

\bibitem{Fermi-note}
J.\ Orear, 
Cornell University Laboratory for Nuclear Studies Preprint 
No.\ CLNS 82/511 (unpublished).  

\bibitem{sterile-nu} 
P.\ Langacker, 
University of Pennsylvania Report No. UPR 0401T (1989);
R.\ Barbieri and A.\ Dolgov, 
Nucl.\ Phys.\ B {\bf 349}, 743 (1991);
K.\ Enqvist, K.\ Kainulainen, and J.\ Maalampi,  
Phys.\ Lett.\ B {\bf 249}, 531 (1990);
M.\ J.\ Thomson and B.\ H.\ J.\ McKellar, 
Phys.\ Lett. B {\bf 259}, 113 (1991);
V.\ Barger {\it et al.}, 
Phys.\ Rev.\ D {\bf 43}, 1759 (1991);
P.\ Langacker and J.\ Liu, 
Phys.\ Rev.\ D {\bf 46}, 4140 (1992).
P.\ I.\ Krastev, S.\ T.\ Petcov, and L.\ Qiuyu
Phys.\ Rev.\ D {\bf 54} 7057 (1996).

\bibitem{Hata-etal-95}
N.\ Hata {\it et al.},
Phys.\ Rev.\ Lett.\ {\bf 75}, 3977 (1995).

\bibitem{Bahcall-Ulmer}
J.\ N.\ Bahcall and A.\ Ulmer,
Phys.\ Rev.\ D {\bf 53}, 4202 (1996). 

\bibitem{Krastev-Petcov}
P.\ I.\ Krastev and S.\ T.\ Petcov,
Phys.\ Rev.\ D {\bf 53} 1665 (1996).

\bibitem{Cumming-Haxton}
A.\ Cumming and W.\ C.\ Haxton,
Phys.\ Rev.\ Lett.\ {\bf 77}, 4286 (1996).


\end{thebibliography}
\end{document}